\documentclass[smallextended]{svjour3}       
\smartqed  
\usepackage{amsmath}
\usepackage{graphicx}
\usepackage{caption}
 \newcommand{\bold}[1]{{#1}}

\usepackage{subfigure}
\usepackage{float}
\usepackage{natbib}
\usepackage{xstring}
\usepackage{xcolor}
\definecolor{red}{cmyk}{0,0,0,1}
\definecolor{blue}{cmyk}{0,0,0,1}
\definecolor{black}{gray}{0}

\begin{document}
\title{The orbital evolution of resonant chains of exoplanets incorporating circularisation
produced by tidal interaction with the central star
with application to the  HD 158259  and EPIC 245950175 systems }
 \subtitle{}

\titlerunning{ The evolution of resonant chains of exoplanets under orbital circularisation
}        

\author{
        J. C. B. Papaloizou }


\institute{ 
       J. C. B. Papaloizou  \at
DAMTP, Centre for Mathematical Sciences, University of Cambridge,
Wilberforce Road, Cambridge CB3 0WA, United Kingdom\\
\email{ J.C.B.Papaloizou@damtp.cam.ac.uk} }

\date{Received: date / Accepted: date}

\maketitle

\begin{abstract}
{We study  orbital evolution of multi-planet systems that form a resonant chain, with nearest neighbours
close to first order commensurabilities, incorporating orbital circularisation produced by tidal interaction with the
central star. We develop a semi-analytic model applicable when the relative proximities to commensurability,
though small,  are large  compared to $\epsilon^{2/3},$ with $\epsilon$
 being a measure of the characteristic planet  to central star mass ratio.
This enables determination of forced
 eccentricities as well as which resonant angles enter libration. When there are no active linked three body Laplace resonances,
 the rate of evolution of the semi-major axes may also be determined.
 We perform numerical simulations of the  HD 158259  and EPIC 245950175 systems
 finding that the semi-analytic approach works well in the former case but not so well in the latter
 case on account of the  effects of three active three body Laplace resonances which persist during the
 evolution. For both systems we  estimate that if the tidal parameter, $Q',$  significantly exceeds $1000,$ tidal effects are unlikely
 to have influenced period ratios significantly since formation. On the other hand if $Q' < \sim 100$
 tidal effects may have produced significant changes including the formation of  three body Laplace resonances
 in the case of  the EPIC 245950175 system.  }

\end{abstract}



\section{Introduction}\label{sec1}
 Hot superEarths or mini-Neptunes with masses in the range $(1 - 20)M_{\oplus},$ orbiting very close to their host stars, 
 have been  discovered  by the Kepler mission   \citep{Batalha2013}.
Many of these are  within compact   systems  containing  pairs that are  close to first order commensurabilities
with some systems comprising or containing a resonant chain with several members. 
Well known examples are Kepler 223 \citep[eg.][]{Lissauer2011}
 and TRAPPIST 1 \citep{Luger2017}.
 
The formation of such systems  readily occurs in scenarios involving orbital migration
\citep[eg.][]{Ward1997, Papaloizou2005, Cresswell2006, Terquem2007, Baruteau2014}.
Although this does not have to  have been extensive. Moreover such chains can be set up,  starting from regions 
close by in phase space, through dissipative effects leading to orbital circularisation, 
during or slightly after the formation process, alone  \citep[see][]{Papaloizou2015, MacDonald2018}. 

An understanding of the post formation evolution is important in order to be able to connect parameters in 
observed systems to conditions just after formation. In general, ubiquitous  migration scenarios
require up to $95\%$ of such systems to be disrupted \citep[eg.][]{Izidoro2017}. Furthermore the period ratios
in systems with close commensurabilities can evolve significantly \citep[eg.][]{Papaloizou2011, Batygin2012},
 and three body Laplace resonances can be set up,
as a  result of orbital circularisation induced by the central star acting on a long time scale \citep{Papaloizou2015},
rather than by  processes operating during formation. In this situation tidal dissipation in the planetary interiors may be significant
for assessing habitability \citep[eg.][]{Papaloizou2018}.

In  this paper  paper we
study the evolution of systems comprising a resonant chain under the  action of orbital
circularisation induced by tidal interaction.  We develop a simple semi-analytic approach, 
as well as   perform numerical simulations, 
making  particular applications  to  the  HD 158259  and EPIC 245950175 systems.

The plan of this paper is as follows. We begin by giving the
basic equations governing a planetary system incorporating orbital circularisation  due to the central star
in Section \ref{sec2}. We  then move on to the development of a simple semi-analytic model in Sections \ref{sec3} and \ref{sec4},
detailing  the approximation scheme used in Sections \ref{sec6}  - \ref{sec9}. 
Using this model the forced eccentricity producing response  is found in Section \ref{sec10} with the potential significance of three body 
Laplace resonances highlighted in Section \ref{sec11}. Conditions for resonance angles to librate, as well as the location of
their centres of libration,  are given in Sections \ref{sec12} and \ref{sec13} with expressions for the rate of change of the
semi-major axes given in Section \ref{sec14}.

Numerical simulations of the  HD 158259  and EPIC 245950175 systems. are  presented in Sections. \ref{Numerics} - \ref{sec21}.
It is found that the semi-analytic model works well in the former case but not so well in the latter on account of the
presence of linked Laplace resonances. We use our results to estimate the rate of evolution of system parameters 
and the dependence on the tidal parameter, $Q'.$ Extrapolation enables us to assess the potential role of
tidal effects in determining the parameters currently observed in these systems.
Finally in Section \ref{Disc} we summarise and discuss our results.

\section{Basic equations governing a planetary system incorporating orbital circularisation  due to the
central star }\label{sec2}
We begin by
considering  a system of $N$ planets and a central star moving in the same plane and interacting gravitationally.
The equations of motion are
\begin{equation}
{d^2 {\bf r}_j\over dt^2} = -{GM{\bf r}_j\over |{\bf r}_j|^3}
-\sum_{k=1\ne j}^N {Gm_k  \left({\bf r}_j-{\bf r}_k \right) \over |{\bf
    r}_j-{\bf r}_k |^3} -{\bf \Gamma} +{\bf \Gamma}_{j} 
    \; ,
\label{emot}
\end{equation}

\noindent  where $M$,  $m_j,$  and ${\bf r}_j$ denote the mass of
the central star,  the mass of planet~$j,$   and the position vector of planet
$j,$ respectively.  The acceleration of the coordinate system based on
the central star (indirect term) is
\begin{equation}
{\bf \Gamma}= \sum_{j=1}^N {Gm_j{\bf r}_{j} \over |{\bf r}_{j}|^3}
\label{indt}
\end{equation}
\textcolor{black}{\bold{and  ${\bf \Gamma}_{j}$   is a frictional damping force that accounts
for orbital circularisation (see below)}.}

\subsection {Orbital circularisation due to tides from the central star}
The circularisation timescale due to tidal interaction with the star
is given by   Goldreich~\& Soter (1966) as
\begin{equation}
t_{e,j} 
=7.63\times10^5\left(\frac{a_j}{0.05au}\right)^{13/2}\left(\frac{M_{\odot}}{M}\right)^{3/2}\left(\frac{M_{\oplus}}{m_{j}}\right)^{2/3}\left(\frac{\rho_{j}}{\rho_{\oplus}}\right)^{5/3}Q' \hspace{2mm}  y.
\label{teccsn}
\end{equation}

\noindent where 
$a_j$ and  
$\rho_{j}$ are the semi-major axis  and the  mean density of the planet. The quantity
$Q'= 3Q/(2k_2),$ where $Q$ is the tidal dissipation function and $k_2$
is the Love number.  
 The values of these  tidal parameters  applicable to  exoplanets  are   unknown. 
 However, for solar system planets in the terrestrial mass
range, Goldreich \& Soter (1966)  estimate  $Q$ to be in the range
10--500 and $k_2 \sim 0.3$, leading to to $Q'$ in the range
50--2500. 

\noindent Orbital circularisation due to  tidal interaction with  the central star
 is dealt with through the addition of a  frictional  damping force taking the form 
 \citep[see eg.][]{Papaloizou2011}
\begin{equation}
{\bf \Gamma}_{j} =
 - \frac{2}{|{\bf r}_j|^2 t_{e,j}} \left( \frac{d {\bf r}_j}{dt} \cdot
{\bf r}_j \right) {\bf r}_j 
\label{Gammai}
\end{equation}

\section{Semi-analytic model for a planetary system consisting of a resonant chain  
undergoing   circularisation}\label{sec3}

We  develop a  model of a system of $N$
planets   undergoing orbital evolution  incorporating the effect of orbital  circularisation
as a result of tidal interaction with the central star.
Torques inducing orbital migration of individual planets may also be included.
However, this aspect will not be explored in detail in this paper.

The planets are  assumed to interact  gravitationally  only with  their inner and outer
neighbours (determined by the value of the semi-major axis).
Equations determining the evolution are obtained  by firstly neglecting
dissipative effects, which are assumed to be small,
 so that the system is governed by a Hamiltonian.
 The effect of dissipative phenomena such as orbital circularisation
 is then added in the simplest manner 
 \citep[see e.g.][]{Papaloizou2015, Papaloizou2018}.

The planets  are assumed to be close enough to  first order resonances
with neighbours so
  that only the resonance angles associated with them need to be retained
  in the Hamiltonian that governs the motion in the absence of dissipative effects
  which we now go on to consider.

\subsection{Hamiltonian formulation}\label{sec4}
We begin by   specifying  the coordinates used before developing the form of the Hamiltonian.
\subsubsection{Coordinates adopted}\label{sec5}
We adopt Jacobi coordinates  \citep[eg.][]{Sinclair1975} for which the radius vector
 of  planet $j,$  ${\bf r }_j,$ is measured relative
to the  centre of mass of  the system comprised of $M$ and  all other  planets
interior to  $j,$ for  $j=1,2, 3, ..., N.$ Here  $j=1$ corresponds to the
innermost planet and $j=N$ to the outermost planet.
 
 \subsubsection {Form of the Hamiltonian }\label{Hamilsec}
The  Hamiltonian for the system governed by (\ref{emot}) with orbital  circularisation absent can be written,  correct to second order
in the planetary masses,  in the form
\begin{eqnarray} H & = &  \sum_{j=1}^N \left({1\over 2}  m_j | \dot {\bf r}_j|^2
- {GM_{j}m_j\over  | {\bf r}_j|} \right)   \nonumber \\
& - &\sum_{j=1}^{N-1}\sum_{k=j+1}^NGm_{j}m_k
\left({1 \over  | {\bf r}_{jk}|}  -  { {\bf r}_j\cdot {\bf r}_k
\over  | {\bf r}_{k}|^3}\right).
\end{eqnarray}
Here $M_{j}=M+m_j $ and
$ {\bf r}_{jk}= {\bf r}_{j}- {\bf r}_{k}.$

Assuming,  the planetary system is strictly coplanar, the equations governing the  motion  
 about a dominant central mass,
 may be written in the form
\citep[see, e.g.][]{ Papaloizou2011}

\begin{eqnarray}
\dot E_j &=& -n_j\frac{\partial H}{\partial \lambda_j}\label{eqnmo1}\\
\dot L_j &=& -\left(\frac{\partial H}{\partial \lambda_j}+\frac{\partial H}{\partial \varpi_j}\right)\\
\dot \lambda_j &=& \frac{\partial H}{\partial L_j} + n_j \frac{\partial H}{\partial E_j}\\
\dot \varpi_j &=& \frac{\partial H}{\partial L_j}.\label{eqnmo4}
\end{eqnarray}

\noindent Here  and in what  follows unless stated otherwise, 
 $m_j$ is replaced by he reduced mass  so that $m_j  \rightarrow m_{j}M/(M+m_{j}).$ 
The orbital  angular momentum of  planet  $j$ 
is $L_j$ and the
orbital  energy is $E_j.$
 The  mean longitude of planet $j$ is $\lambda_j = n_j (t-t_{0j}) +\varpi_j ,$
 with  $n_j  = \sqrt{GM_{j}/a_j^3}= 2\pi/P_j $ being  the  mean motion,  and
$t_{0j}$ denoting the time of periastron passage.  The semi-major axis and orbital period   
 of planet  $j$ are  $a_j$  and $P_j.$   
 The longitude of periastron is $\varpi_j.$
 The quantities $\lambda_j,$ $\varpi_j,$ $L_j$ and $E_j$ can be used  as to describe the dynamical
 system described above.

\noindent For motion around a central point  mass $M$ the angular momentum and energy  of planet, $j,$
are related to its semi-major axis and eccentricity through the relations
\begin{eqnarray}
     L_j &=&  m_{j}\sqrt{GM_{j}a_i(1-e_j^2)}, \\
     E_j &=& -{{GM_{j}m_{j}}\over{2a_j}},
\end{eqnarray}
where $e_j$  the eccentricity of planet $j.$ 
By making use of these relations  we  may adopt 
$\lambda_j,$ $\varpi_j,$ $a_j$ or equivalently $n_j,$  and $e_j$ as dynamical variables. 
 We comment that the difference between  taking $m_j$
to be the reduced mass rather than  the actual mass  of planet $j$ when evaluating $M_j$  in the  expressions for
 $L_j$ and $E_j$
is third order in the typical planet to star mass ratio and thus it may be neglected.
The equations we  ultimately  use  turn out to be effectively equivalent to  those obtained assuming the central mass is fixed. 
 The Hamiltonian may quite generally
 be expanded in a Fourier series
involving linear combinations of the $2N-1$   angular differences
$\varpi_j -\varpi_1,$ $j=2, 3, ...,  N$ and
$\lambda_j - \varpi_j,  j=1,2,3, ..., N. $

\subsubsection{Commensurabilities and averaging}\label{sec6}
 Here we suppose that the important interactions are through
the effects of the $N-1$  first order commensurabilities, $p_j+1: p_j, $ with $p_j$ being a positive integer,  
associated with the planets  with masses  $m_j$ and $m_{j+1},$ for $j=1,2,...N-1,$ respectively.
Corresponding to  this situation,  we expect that any of the $2(N-1)$  angles
$\Phi_{j+1,j,1} = (p_j+1)\lambda_{j+1}-p_j\lambda_j-\varpi_j, $ 
$\Phi_{j+1,j,2} = (p_{j}+1)\lambda_{j+1}-p_j\lambda_{j}-\varpi_{j+1},$ $j=1,2,3, ...,  N-1,$
will  be slowly varying.
 Following  standard practice
 \citep[see, e.g.][]{ Papaloizou2015, Papaloizou2018},
high frequency terms in the Hamiltonian  are not expected to be of comparable importance and
are  accordingly averaged out.
In this way,  only terms in the Fourier expansion involving  linear
combinations of $\Phi_{j+1,j,1},$  and $\Phi_{j+1,j,2},$ for $j=1,2,3, ..., N-1,$
as argument are  retained.

The eccentricity is assumed to be small  such that
 terms that are higher order than first  in the eccentricities can  be  neglected.
The  Hamiltonian   may  then be written
in the form
\begin{equation} H=\sum_{k=1}^{N}E_k +  \sum_{k=1}^{N-1}  H _{k,k+1} ,\label{Hamil0} \end{equation}
 where
\begin{equation} \hspace{-1mm} H _{k, k+1}= -\frac{Gm_km_{k+1}}{a_{k+1}}\left[  e_{k+1}C_{k,k+1}\cos \Phi_{k+1,k,2}
- e_kD_{k,k+1}\cos\Phi_{k+1, k, 1} \right] \label{Hamil} \end{equation}
with
\begin{eqnarray} \hspace{-6mm}C _{k,k+1} & = & {1 \over 2}\left(   \alpha{db^{m}_{1/2}(\alpha)\over d\alpha} 
 +(2m+1)b^{m}_{1/2}(\alpha)
-(2m+2)\alpha\delta_{m,1} \right),   \label{Hamil1} \\
\hspace{-6mm}D _{k,k+1} & =& {1 \over 2}\left(   \alpha{db^{m+1}_{1/2}(\alpha)\over d\alpha}
 +2(m+1)b^{m+1}_{1/2}(\alpha)  \right) . \label{Hamil2} 
\end{eqnarray}
Here the integer $m=p_k,$ 
 $b^{m}_{1/2}(\alpha)$ denotes  the usual Laplace coefficient
 \citep[eg.][]{Brouwer1961, Murray1999}
with the argument $\alpha = a_k/a_{k+1}.$ 

 Using equations~(\ref{eqnmo1})--(\ref{eqnmo4})
together with equation~(\ref{Hamil0})  we  obtain the equations of motion in the form

\begin{align}
&\hspace{-.3cm}\frac{d e_j}{dt}=\frac{n_j}{M_j}\left[ m_{j+1}  \frac{a_j}{a_{j+1}}
D_{j,j+1}\sin\Phi_{j+1, j, 1}                                
- m_{j-1} C_{j-1,j}\sin \Phi_{j,j-1,2}\right]  , \label{eqntid10a}\\
\!\!\!\!\!\!\!\!\!\!\!
&\hspace{-.3cm} \frac{d n_j}{dt} =
\frac{3(p_{j-1}+1)n_j^2m_{j-1}}{M_j}\biggl[C_{j-1,j}e_j\sin \Phi_{j,j-1,2}-D_{j-1,j }e_{j-1}\sin \Phi_{j,j-1,1} \biggr]\nonumber\\
   &\hspace{-.4cm}-
\frac{3p_j n_j^2m_{j+1}a_j}{M_ja_{j+1}}\biggl[C_{j,j+1}e_{j+1}\sin\Phi_{j+1,j,2}-D_{j,j+1}e_j\sin\Phi_{j+1,j,1} \biggr]
 \label{eqntid20a} \\
\!\!\!\!\!\!\!\!\!\!
&\vspace{0.4cm}\nonumber\\
\!\!\!\!\!\!\!\!\!\!
&\hspace{-.2cm}
{\rm for {\hspace{2mm} }}j = 1,2,3,4, ..., N\hspace{6mm} {\rm and}\nonumber \\
\!\!\!\!\!\!\!\!\!\!
&\vspace{0.5cm}\nonumber\\
&\hspace{-.3cm}\frac{d  \Phi_{j+1,j,1}}{dt} = (p_j+1)n_{j+1}-p_jn_j\nonumber\\
&\hspace{-.3cm}-\frac{n_j}{e_j}\left[ \frac{m_{j-1}}{M_j} C_{j-1,j}\cos \Phi_{j,j-1,2}-\frac{m_{j+1}a_{j}}{M_{j} a_{j+1}}D_{j,j+1}\cos \Phi_{{j+1,j,1}}\right]   ,
\hspace{2mm}{\rm with}  \label{eqntid30}\\
&\hspace{-.3cm}\frac{d  \Phi_{j+1,j,2}}{dt } = (p_j+1)n_{j+1}-p_jn_j\nonumber\\
&\hspace{-.3cm}-\frac{n_{j+1}}{e_{j+1}}\left[ \frac{m_{j}}{M_{j+1}} C_{j,j+1}\cos \Phi_{j+1,j,2}-\frac{m_{j+2}a_{j+1}}{M_{j+1} a_{j+2}}D_{j+1,j+2}\cos \Phi_{{j+2,j+1,1}}\right] .
 \label{eqntid40}\\
 \!\!\!\!\!\!\!\!\!\!
 \!\!\!\!\!\!\!\!\!\!
& \vspace{4mm}\nonumber\\
&\hspace{-.2cm}
{\rm for {\hspace{2mm} }}j = 1,2,3, ... ,N-1 \hspace{6mm} \nonumber \\
\nonumber
\end{align}





\subsubsection{ Incorporation of dissipative effects}\label{sec7}
The effect of   orbital circularisation  due to   tidal interaction with the central star
may be included by  adding the eccentricity damping term
$- {e_j}/{t_{e,j}}$ to equation (\ref {eqntid10a})
and the term corresponding to the induced energy dissipation
${3n_je_j^2}/{t_{e,j}}$ to equation(\ref{eqntid20a}).
{\bold{ We remark that the latter term  is  second order in eccentricity whereas only first order terms were considered
in Section \ref{sec6}. However, that corresponds to
 the lowest order at which changes to  the total energy of the system occur.
That dissipative effects can be incorporated in this way  without adding in higher order non dissipative effects
is a common assumption in semi-analytic treatments of the type undertaken below. 
These are later checked  with numerical simulations.}}

We  remark that the effect of torques leading to orbital migration
   can be incorporated by adding an additional term
${n_j}/{t_{mig,j}}$ to equation(\ref{eqntid20a}), where $t_{mig,j}$  defines a migration time
of planet $j.$ 
{\bold{It is well known that such torques can lead to the setting up of commensurabilities
through convergent migration and to resonant chains when many planets are involved
\citep[see eg. ][]{Baruteau2014, Papaloizou2005, Papaloizou2018}}.
However, we shall not discuss the potential role of such torques
further in this paper.
Incorporating \bold{orbital circularisation as indicated above}} 
 equations (\ref{eqntid10a}) and (\ref{eqntid20a}) respectively become

\begin{align}
&\hspace{-.3cm}\frac{d e_j}{dt}=\frac{n_j}{M_j}\left[ m_{j+1}  \frac{a_j}{a_{j+1}}
D_{j,j+1}\sin\Phi_{j+1, j, 1}                                
- m_{j-1} C_{j-1,j}\sin \Phi_{j,j-1,2}\right]  - \frac{e_j}{t_{e,j}}, \label{eqntid10}\\
&\hspace{0.7cm} {\rm and} \nonumber\\
&\vspace{1cm}\nonumber\\
&\hspace{-.3cm} \frac{d n_j}{dt} =
\frac{3(p_{j-1}+1)n_j^2m_{j-1}}{M_j}\biggl[C_{j-1,j}e_j\sin \Phi_{j,j-1,2}-D_{j-1,j }e_{j-1}\sin \Phi_{j,j-1,1} \biggr]\nonumber\\
   &\hspace{-.4cm}-
\frac{3p_j n_j^2m_{j+1}a_j}{M_ja_{j+1}}\biggl[C_{j,j+1}e_{j+1}\sin\Phi_{j+1,j,2}-D_{j,j+1}e_j\sin\Phi_{j+1,j,1} \biggr]\nonumber\\
&\hspace{-.3cm}+\frac{3n_je_j^2}{t_{e,j}}  
, \label{eqntid20} \\
&\vspace{0.4cm}\nonumber\\
&\hspace{-.2cm}{\rm for {\hspace{2mm} }}j = 1,2,3,4, ..., N .\nonumber 
\end{align}
\noindent We remark that terms on the right hand sides of the above equations for  which $j$  takes on a value such that  a factor  $m_0$ or $m_{N+1}$ is implied 
are to be omitted  or one may set $m_0 = m_{N+1} =0.$ From now on we shall adopt the latter convention.

\subsection{Development of an approximation scheme applicable when the semi-major axis variations are small} \label{sec8}
We shall consider the situation when the system is such that the commensurabilities  are significant but departures 
from exact commensurability are large enough
that variations in the semi-major axes can be neglected when calculating  forced eccentricities.
This corresponds to calculating the response, or epicyclic motion, induced by interaction  of a planet with its neighbours
assuming that these are on fixed circular orbits.

We begin by defining a new set of variables $(x_j , y_j)$	such that																			

\noindent $x_j= e_j\sin\Phi_{j+1,j,1}$ and  $y_j= e_j\cos\Phi_{j+1,j,1}$ for $j = 1,2,...N-1,$ with

\noindent $x_N= e_N\sin\Phi_{N,N-1,2}$ and  $y_N= e_N\cos\Phi_{N,N-1,2},$
 and a new variable 
 
\noindent $z_j=1/n_j -1/n_{j,0}, $ where $n_{j,0}$ is a constant reference value of $n_{j}.$
 Substituting these into equations (\ref{eqntid30}) - (\ref{eqntid20}) we obtain
 
\begin{align}
&\hspace{-0.7cm}\frac{d  x_j}{dt}=((p_j+1)n_{j+1}-p_jn_j)y_j \nonumber\\
&\hspace{-0.8cm}+\frac{n_j}{M_j}\left[ m_{j+1}  \frac{a_j}{a_{j+1}}
D_{j,j+1}                             
- m_{j-1} C_{j-1,j}\cos \beta_{j}\right]  - \frac{x_j}{t_{e,j}}, \label{eqntid1}\\
\!\!\!\!\!\!\!\!\!\!\!
&\hspace{-0.7cm}\frac{d y_j}{dt}= -((p_j+1)n_{j+1}-p_jn_j)x_j+
\frac{n_j}{M_j}                             
 m_{j-1} C_{j-1,j}\sin \beta_{j} - \frac{y_j}{t_{e,j}}, \label{eqn000}\\
\!\!\!\!\!\!\!\!\!\!
&\vspace{0.4cm}\nonumber\\
\!\!\!\!\!\!\!\!\!\!
&\hspace{-.7cm}
{\rm for {\hspace{2mm} }}j = 1,2,3,4, ..., N -1 \hspace{6mm} {\rm with}\nonumber \\
\!\!\!\!\!\!\!\!\!\!
&\vspace{0.5cm}\nonumber\\
\frac{d x_N}{dt}&=((p_{N-1}+1)n_{N}-p_{N-1}n_{N-1})y_N 
-\frac{n_{N}}{M_{N}}       
 m_{N-1} C_{N-1,N}  - \frac{x_N}{t_{e,N}}, \label{eqntid11}\\
\!\!\!\!\!\!\!\!\!\!\!
\frac{d y_N}{dt}&= -((p_{N-1}+1)n_{N}-p_{N-1}n_{N-1})x_{N}
- \frac{y_N}{t_{e,N}}, \label{eqn0001}\\
\!\!\!\!\!\!\!\!\!\!
&\vspace{0.4cm}\nonumber\\
\!\!\!\!\!\!\!\!\!\!
&\hspace{-0.8cm}
 {\hspace{2mm} } {\rm and}\nonumber \\
\!\!\!\!\!\!\!\!\!\!
&\vspace{0.5cm}\nonumber\\
&\hspace{-0.7cm} \frac{d z_j}{dt} =
-\frac{3(p_{j-1}+1)m_{j-1}}{M_j}\biggl[C_{j-1,j}(x_j\cos \beta_{j}-y_j\sin\beta_j)-D_{j-1,j }x_{j-1} \biggr]\nonumber\\
   &\hspace{-.8cm}+
\frac{3p_j m_{j+1}a_j}{M_ja_{j+1}}\biggl[C_{j,j+1}(x_{j+1}\cos\beta_{j+1}-y_{j+1}\sin\beta_{j+1})     -D_{j,j+1}x_j \biggr]\nonumber\\
&\hspace{-.8cm}-\frac{3(x_j^2+y_j^2)}{n_jt_{e,j}}  
 , \label{eqntid2} \\
&\vspace{0.5cm}\nonumber\\
&\hspace{-.7cm}
{\rm for {\hspace{2mm} }}j = 1,2,3,4, ..., N .\hspace{6mm} \nonumber \\
\nonumber
\end{align}
Here we have set   $\beta_j=  \Phi_{j+1,j,1} - \Phi_{j, j -1,2}$ for  $j  = 2,3, ..., N-1.$
The latter definition is not applicable for $j=1$ or $j=N.$
In practice we find it convenient and consistent with the equations we use 
to  adopt the convention  of setting  $\beta_1 =\beta_N =0$ along with $m_0=m_{N+1}=0$ where
the notation implies these appear.
\subsubsection{Scaled variables and ordering scheme}\label{sec9}
We now set up  an ordering scheme depending on two small parameters.
The first, $\epsilon$ is a characteristic mass ratio $m_j/M$
 (we assume this is the same order independent of $j$).
 The second, $\lambda$ is such that $\epsilon^{2/3}/\lambda$ measures 
 the departures from the  first order commensurabilities  associated with the resonant angles
 in the development in Section \ref{sec6}. In order that the deviation from commensurability
 be small, $\lambda$  may be small but $ > O(\epsilon^{2/3}),$ for example a possibility
 is that $\lambda$ is $O(\epsilon^{1/3}).$   For simplicity we shall  suppose that 
 a single pair  of parameters applies to all the planets in. a system rather than attempt
to taylor a system to individual planets.
  
In addition we consider solutions for which $n_j$ is close to some value $n_{j,0}$
 associated with  a base  state indicated with a subscript, $0,$  and define  scaled variables
indicated with a $ \tilde\\  $  over them  such that
\begin{align}
 &\hspace{-5.9cm} x_j=\tilde x_j\epsilon^{1/3}\lambda, \label{scal1}\\ 
  &\hspace{-5.9cm}y_j=\tilde y_j\epsilon^{1/3}\lambda.\label{scal2} \hspace{3mm} {\rm and} \hspace{3mm}\\ 
&\hspace{-5.9cm} (p_j+1)n_{{j+1}}-p_jn_{j} = {\tilde \omega}_{j+1,j}\epsilon^{2/3}/\lambda. \label{scal3}
\end{align}
Along with this, with reference to the base state we define
\begin{align}
\hspace{-5.4cm} (p_j+1)n_{{j+1},0}-p_jn_{j,0} = ({\tilde \omega}_{j+1,j})_0\epsilon^{2/3}/\lambda. \label{scal4}
\end{align}
The intention here is that the scalings are chosen such  the quantities ${\tilde x}$  and  ${\tilde y}$ 
will be of order unity
while ${\tilde \omega_{j+1,j}}$ with be comparable to $n_{j,0}$ 
in magnitude (note that $\epsilon$ and $\lambda$ are assumed to be positive with ${\tilde \omega_{j+1,j}}$
being of either sign). 
In addition we find it convenient to define ${\tilde z_j}$ through 
 \begin{align}
 \hspace{-5.9cm} z_j= \epsilon^{2/3} {\tilde z}_j\label{scal5}
 \end{align}
  and  a scaled time $\tau$ through
  \begin{align}
  \hspace{-5.9cm}  t=\tau \lambda\epsilon^ {-2/3}.\label{scal6}
   \end{align}
Here we expect that the characteristic magnitude of  ${\tilde z}_j$ will be of order $1/n_{j,0}$ 
and we shall see that  $(n_j-n_{j,0})/n_{j,0},$  which gives the characteristic 
magnitude of the relative  amplitude of oscillations of the semi-major axes,
 will be of order $\lambda^2\epsilon^{2/3},$ which from (\ref{scal1}) is characteristically the square of a forced eccentricity.
 
Together with (\ref{scal4}) this implies that the ratio of the relative variation in the semi-major axes
to the characteristic relative deviation from commensurability is of order $\lambda^3.$
When this is small, as is assumed,  fluctuations of the semi-major axes will not affect the closeness to commensurability 
and thus  may be neglected when
calculating the forced eccentricities at the lowest order approximation.

Expressed in terms of the above scaled variables, Equations (\ref{eqntid1}) - (\ref{eqn0001})
 lead to
\begin{align}
\frac{d \tilde x_j}{d\tau}&=
(\tilde\omega_{j+1,j})_0\tilde y_j +
\frac{n_{j,0}}{\epsilon M_j}\left[  m_{j+1} \left( \frac{a_j}{a_{j+1}}
D_{j,j+1} \right)_0                            
- m_{j-1}\left( C_{j-1,j}\right)_0\cos \beta_{j}\right]  \nonumber\\
\!\!\!\!\!\!\!\!\!\!\!
&\hspace{-0.7cm}+O(\lambda^3 )+ O(\lambda^2\epsilon^{2/3}) - \frac{\tilde x_j}{\tilde t_{e,j}}, \label{eqntid1s}\\
\!\!\!\!\!\!\!\!\!\!\!
&\vspace{1cm}\nonumber\\
\frac{d \tilde y_j}{d\tau}&= -(\tilde\omega_{j+1,j})_0\tilde x_j+
 \frac{n_{j,0}}{\epsilon M_j}                           
m_{j-1}\left( C_{j-1,j}\right)_0\sin \beta_{j} \nonumber\\
\!\!\!\!\!\!\!\!\!\!
&\hspace{-0.7cm}+O(\lambda^3 )+ O(\lambda^2\epsilon^{2/3})- \frac{\tilde y_j}{\tilde t_{e,j}}, \label{eqn000s}\\
\!\!\!\!\!\!\!\!\!\!
\!\!\!\!\!\!\!\!\!\!
&\vspace{2cm}\nonumber\\
&\hspace{-.5cm}
{\rm for {\hspace{2mm} }}j = 1,2,3,4, ..., N -1 \hspace{6mm} {\rm with}\nonumber \\
\!\!\!\!\!\!\!\!\!\!
\!\!\!\!\!\!\!\!\!\!
&\vspace{2cm}\nonumber\\
&\hspace{-0.6cm} \frac{d \tilde x_N}{d\tau}=(\tilde\omega_{N,N-1})_0\tilde y_N 
-\frac{n_{N,0}}{\epsilon M_{N}}       
 m_{N-1} \left(C_{N-1,N} \right)_0  \nonumber\\
 &\hspace{-0.7cm} +O(\lambda^3 )+ O(\lambda^2\epsilon^{2/3}) - \frac{\tilde x_N}{\tilde t_{e,N}} \label{eqntid11s}\\
\!\!\!\!\!\!\!\!\!\!\!
&\hspace{-0.6cm}{\rm and}\nonumber\\
&\vspace{1cm}\nonumber\\
&\hspace{-0.6cm} \frac{d\tilde y_N}{d\tau}= -(\tilde \omega_{N,N-1})_0\tilde x_{N}+O(\lambda^3 )
- \frac{\tilde y_N}{\tilde t_{e,N}}. \label{eqn0001s}\\
\nonumber\\
\nonumber
\end{align}
Here  we  assume $t_{e,j}$ 
is constant 
or equivalently  evaluated for the background state  with the subscript $0$ being  dropped
and we have $\tilde t_{e,j} =\epsilon^{2/3}\lambda^{-1} t_{e,j} $ 
with  $O(\lambda^3 )+ O(\lambda^2\epsilon^{2/3})$ indicating that additional omitted terms are either of order $\lambda^3$
or $\lambda^2\epsilon^{2/3}$ compared to those retained. These will subsequently be neglected. 
However, it should be noted that these corrections are derived for the simplified
system governed by (\ref{eqntid30}) - (\ref{eqntid20}) for which high frequency corrections have been dropped.
Such corrections may appear in the analogues of  (\ref{eqntid1s}) - (\ref{eqn0001s}) when the full system is considered
and have larger amplitude than implied by the magnitude of the above corrections.
Notably the simple  model assumes that they can be averaged out. 
We note that the subscript $0$ attached to 
 a bracket as well as a particular quantity indicates
evaluation at the background state with $n_j=~n_{j,0}.$
Following the same procedure in the case of  equation  (\ref{eqntid2}) leads to 
\begin{align}
&\hspace{-0.2cm} \frac{1}{\lambda^2} \frac{d\tilde z_j}{d\tau} =
-\frac{3(p_{j-1}+1)m_{j-1}}{\epsilon M_j}
\biggl[(C_{j-1,j})_0(\tilde x_j\cos \beta_{j}-\tilde y_j\sin\beta_j)
-\left(D_{j-1,j }\right)_0
\tilde x_{j-1}
\biggr] 
 \nonumber\\
   &\hspace{-.2cm}+
\frac{3p_j m_{j+1}}{\epsilon M_j}\left(\frac{a_j}{a_{j+1}}\right)_0\biggl[ (C_{j,j+1})_0(\tilde x_{j+1}\cos\beta_{j+1}-\tilde y_{j+1}\sin\beta_{j+1})  
  -(D_{j,j+1})_0\tilde x_j 
\biggr]  \nonumber\\
&\hspace{-.2cm}-\frac{3(\tilde x_j^2+\tilde y_j^2)}{n_{j,0}\tilde t_{e,j}}  
 , \label{eqntid2s} \\
&\vspace{1cm}\nonumber\\
&\hspace{-0.0cm}
{\rm for {\hspace{2mm} }}j = 1,2,3,4, ..., N .\hspace{6mm} \nonumber \\
\nonumber
\end{align}
 Importantly for our application, we remark that equation (\ref{eqntid2s})
indicates that the amplitude of oscillations in ${\tilde z}_j$ is reduced by a factor of $\lambda^2$
 as compared to the magnitude  of   ${\tilde x}_j.$  Using the scaling relations 
 (\ref{scal1}) - (\ref{scal5}) and, given that ${\tilde x}_j$ is of order unity, this implies that the relative amplitude of
  semi-major axes oscillations is  $\sim \epsilon^{2/3}\lambda^2\sim e_j^2$
as was indicated above (see discussion below equation (\ref{scal6})).
 Accordingly as was also  indicated there, 
 this enables us to adopt a strategy
of determining the evolution of the eccentricities assuming that the semi-major axes do not change
and then using the results to determine the slow rates of change of the semi-major axes.
\subsection{Finding the  forced eccentricities}\label{sec10}
As the first step in determining  the evolution of the eccentricities, we note that
from   
 equations    (\ref{eqntid20}) and (\ref{eqntid30}) we find that 
 \begin{align}
 &\hspace{-4.5cm} \frac{d\beta_j}{dt}=\frac{d  \Phi_{j+1,j,1}}{dt} -\frac{d  \Phi_{j,j-1,2}}{dt}\nonumber\\
 &\hspace{-4.5cm} = (p_j+1)n_{j+1}- (p_{j}+p_{j-1}+1)n_{j}+p_{j-1}n_{j-1}\label{Laplacec}\\
 &\vspace{1cm}\nonumber\\
 &\hspace{-4.5cm} {\rm for} \hspace{2mm} j=2,3, ..., N-1 \nonumber
 \end{align}
 or in terms of scaled variables
  \begin{align}
 &\hspace{-3.6cm} \frac{d\beta_j}{d\tau}=  \tilde\omega_{j+1,j} -\tilde\omega_{j,j-1}=
 (\tilde\omega_{j+1,j})_0 -(\tilde\omega_{j,j-1})_0 +O(\lambda^3),
 \label{Laplacec1}
 \end{align}
where $O(\lambda^3)$ indicates corrections due to the variations of the mean motions
that are small when $\lambda$ is small and that accordingly will be neglected from now on, though we shall bear in mind
that in addition, rapidly oscillating corrections have been averaged out in the model,  and  take care about noting their presence.
\subsubsection{Condition for a Laplace resonance}\label{sec11}
Notably the condition for the right hand side of equation (\ref{Laplacec}) to vanish corresponds to the
condition for a Laplace resonance. It is important to note that if it is satisfied for the background state then
under the approximation that the variation of the semi-major axes is neglected we find that $\beta_j$ is a constant
that is not determined in this approximation scheme. In reality it should be regarded as slowly varying,
with the variation being determined at a higher order of approximation.
This means that the description will be incomplete at the lowest order approximation used here
when there is a strict Laplace resonance (for more details see below).
 \subsubsection{ Determining forced eccentricities}\label{sec12}
 We now determine the epicyclic response by solving
 equations (\ref{eqntid1s}) - (\ref{eqn0001s}) and equation (\ref{Laplacec1}) with corrections
 $O(\lambda^3)+O(\lambda^2\epsilon^{2/3})$ neglected.
 It can be seen that this
 this amounts to solving a linear forced harmonic oscillator problem.
 In doing this we find the solution assuming that transients have decayed  which will have happened
 on the circularisation time, $t_{e,j}.$ The amplitudes $\sqrt{\tilde x_j^2+\tilde y_j^2}$ correspond to the forced eccentricities of the planets induced
 by the  perturbations of their neighbours assumed to be on circular orbits for this purpose.

  It is easy to. show that the solution described above can be written in the form

  \begin{align}
 \hspace{-0.2cm}&{\tilde x_{j}} = -\frac{{{\cal \alpha}_j (\cos\beta_j}/t_{e,j}  
 -(\tilde\omega_{j,j-1})_0\sin\beta_j)}{(\tilde\omega_{j,j-1})_0^2+1/\tilde t_{e,j}^{2}} 
 + \frac{\gamma_j/\tilde t_{e,j}}
 {( (\tilde\omega_{j+1,j})_0^2+1/\tilde t_{e,j}^2)} \hspace{3mm}{\rm and }\hspace{2mm}  \label{eccsol1}\\
 \hspace{-0.2cm}
 &{\tilde y_{j}} = \frac{{{\cal \alpha}_j ((1/ \tilde t_{e,j}) \sin\beta_j} 
   +(\tilde\omega_{j,j-1})_0\cos\beta_j)}{({\tilde\omega_{j,j-1})_0^2+1/\tilde t_{e,j}^{2}}} 
 - \frac{\gamma_j(\tilde \omega_{j+1,j})_0}
 { ((\tilde\omega_{j+1,j})_0^2+1/\tilde t_{e,j}^2)},
    \label{eccsol2}        
 \end{align}
 \vspace{-0.3cm}
 \begin{align}
&\hspace{-2.9cm} {\rm for} \hspace{2mm} j= 1,2,3, ..., N-1,\hspace{2mm} {\rm where} \hspace{3mm} {\cal \alpha }_j= \frac{n_{j,0}}{\epsilon M_j}m_{j-1} \left(C_{j-1,j}\right)_0\nonumber\\
&\hspace{-2.9cm}{\rm and}\hspace{3mm}
 {\cal \gamma }_j = \frac{n_{j,0}}{\epsilon M_j}  m_{j+1}  \left(\frac{a_j}{a_{j+1}}D_{j,j+1}\right)_0.\nonumber
 \end{align}
 We remark that with our notation convention $\alpha_1=\gamma_N =0.$
 In addition
 \begin{align}
&\hspace{-.4cm} \tilde x_{N} = - \frac{\alpha_{N}}
 { \tilde t_{e,N}((\tilde\omega_{N,N-1})_0^2+1/\tilde t_{e,N}^2)},\hspace{2mm}{\rm and}\hspace{2mm}
\tilde y_{N} =   \frac{\alpha_{N}(\tilde\omega_{N,N-1})_0}
 { ((\tilde\omega_{N,N-1})_0^2+1/\tilde t_{e,N}^2)}.\label{XNYN}
 \end{align}
\subsubsection{Conditions for libration}\label{sec13}
From equations (\ref{eccsol1}) and (\ref{eccsol2}) we find that $(\tilde x_j, \tilde y_j)$ lies on the circle
\begin{align}
&\hspace{-2.4cm}\left({\tilde x_{j}}  -\frac{\gamma_j/\tilde t_{e,j}}
 {( (\tilde\omega_{j+1,j})_0^2+1/\tilde t_{e,j}^2)}\right)^2 +
 \left({\tilde y_{j}} +  \frac{\gamma_j(\tilde \omega_{j+1,j})_0} { ((\tilde\omega_{j+1,j})_0^2+1/\tilde t_{e,j}^2)}\right)^2\nonumber\\
&\hspace{-2.4cm}  = \frac{{\cal \alpha}_j ^2}
 { ((\tilde\omega_{j,j-1})_0^2+1/\tilde t_{e,j}^2)}
\end{align}   
{\bold{with centre at 
 \begin{align} 
\hspace{-1.6cm} \left (  \frac{\gamma_j/\tilde t_{e,j}}
 {( (\tilde\omega_{j+1,j})_0^2+1/\tilde t_{e,j}^2)} ,   -  \frac{\gamma_j(\tilde \omega_{j+1,j})_0} { ((\tilde\omega_{j+1,j})_0^2+1/\tilde t_{e,j}^2) }   \right)
{\rm 
 in \  the \  }  ({\tilde x}, {\tilde y})  {\rm  \  plane}.
 \nonumber
 \end{align}}}
\hspace{-2.5mm}  Accordingly, noting that $e_j$ is the cylindrical polar radius and $\pi/2-\Phi_{j+1,j,1} $ is the cylindrical polar angle,
 as $(\tilde x_j,\tilde y_j)$ moves on this 
circle,  the condition for libration of  $\Phi_{j+1,j,1}$  is that the circle does not enclose the origin in the $(\tilde x_j, \tilde y_j)$ plane.
This in turn implies that
\begin{align}
\hspace{-2.5cm}&\frac{\gamma_j^2}
 { ((\tilde\omega_{j+1,j})_0^2+1/\tilde t_{e,j}^2)}
  >\frac{{\cal \alpha}_j ^2}
 { ((\tilde\omega_{j,j-1})_0^2+1/\tilde t_{e,j}^2)}.\label{condlib1}
\end{align}   
One can also see that in the limit of large circularisation times,  {\bold{ for which  we may assume that, 
$|(\tilde\omega_{j,j-1})_0 |\tilde t_{e,j} \gg 1,$
which is the case of interest here, 
the centre of the circle will lie very close to the positive/negative  ${\tilde y}$ axis
according to whether,  $ \gamma_j/(\omega_{j+1,j})_0,$ 
  is negative/positive. 
Corresponding to this the libration  of $\Phi_{j+1,j,1} $}}
will be about zero or $\pi$ according to whether $ \gamma_j/\omega_{j+1,j}$ is negative or positive.

Similarly, by considering  the trajectory of $(\tilde x'_j = \tilde x_j\cos\beta_j -\tilde y_j\sin\beta_j, \tilde y' = \tilde y_j\cos\beta_j +\tilde x_j\sin\beta_j)$
in the $ (\tilde x_j', \tilde y_j')$ plane the condition for the libration of the angle  $\Phi_{j,j-1,2}$
is found to be
\begin{align}
\hspace{-2.6cm}&\frac{\gamma_j^2}
 { ((\tilde\omega_{j+1,j})_0^2+1/\tilde t_{e,j}^2)}
  <\frac{{\cal \alpha}_j ^2}
 { ((\tilde\omega_{j,j-1})_0^2+1/\tilde t_{e,j}^2)}.\label{condlib2}
\end{align} 
In this case  one finds that in the limit of large circularisation times that
 the libration will be about zero or $\pi$ according to whether $ \alpha_j/(\tilde \omega_{j,j-1})_0$ is positive or negative.

The above discussion  indicates that one of $\Phi_{j+1,j,1}$ or $\Phi_{j,j-1,2}$  may librate but not both
An exception occurs when an angle $\beta_j$ is constant. In that case the phase points do not move around  the circles
and are thus fixed corresponding to zero. amplitude libration. From  equation (\ref{Laplacec}), as noted above
 we recall that this special condition corresponds to 
a Laplace resonance for which
\begin{align}
\hspace{-2.1cm}&(p_j+1)n_{j+1}- (p_{j}+p_{j-1}+1)n_{j}+p_{j-1}n_{j-1}=0\label{Lapstrict}
\end{align}
is evaluated for the reference base state.
For the special cases with $j=1$ and $j=N,$ \textcolor{red}{as the terms involving  $\alpha_1$  and  $\gamma_N $ that respectively appear in
the conditions (\ref{condlib1}) and (\ref{condlib2}) are zero, these  imply} that $\Phi_{2,1}$ and  $\Phi_{N,N-1,2}$ 
are librating after transients decay. 
 \subsection{The rate of change of the semi-major axes}\label{sec14}
Substituting the  the eccentricities given by  (\ref{eccsol1}) - (\ref{XNYN}) into equation (\ref{eqntid2s}) and taking a time average, we obtain
an equation from which the mean  rate of change of $\tilde z_{j}$ and  hence the semi-major axes may be found.
Typically the time scale involved is the product of $e_j^{-2}$ and the circularisation time which is expected to be very much longer than
the time scale associated with the oscillation of the angle, $\beta_j,$ justifying taking a time average.
In this way we find
 \begin{align}
&\hspace{-0.3cm} \frac{1}{\lambda^2} \frac{d\tilde z_j}{d\tau} =
\frac{3(p_{j-1}+1)\alpha_{j}}{ n_{j .0}} 
\left(\frac{\alpha_j/\tilde t_{e,j}}{({\tilde\omega_{j,j-1}^2+1/\tilde t^2_{e,j} } )} 
+\frac{D_{j-1,j}}{C_{j-1,j}}\frac{\gamma_{j-1}/\tilde t_{e,j-1}}{({\tilde\omega_{j,j-1}^2+1/\tilde t^2_{e,j-1} } )}  \right)_0\nonumber\\
   &\hspace{-.4cm}
-\frac{3p_j \gamma_{j}}{ n_{j ,0}}\frac{C_{j,j+1}}{D_{j,j+1}}  
\left(
\frac{\alpha_{j+1}/\tilde t_{e,j+1} }{ ({\tilde\omega_{j+1,j}^2+1/\tilde t^2_{e,j+1} } )} 
           +\frac{D_{j,j+1}}{C_{j,j+1}}  \frac{\gamma_{j}/\tilde t_{e,j}}{({\tilde\omega_{j+1,j}^2+1/\tilde t^2_{e,j} } )}\right)_0\nonumber\\
&\hspace{-.4cm}-\frac{3}{n_{j,0}\tilde t_{e,j}}
\left ( \frac{\alpha_j^2}{({\tilde\omega_{j,j-1}^2+1/\tilde t^2_{e,j} } )} + \frac{\gamma^2_{j}}{({\tilde\omega_{j+1,j}^2+1/\tilde t^2_{e,j} } )}  \right  )_0 
    , \label{eqntid2snew} 
\end{align} 
for $j=1,2,....N.$

It is important to note that when there is a Laplace resonance $\beta_j$ is an undetermined constant within this approximation 
scheme and so the terms involving it cannot be averaged out. In reality its behaviour is determined by terms that have been neglected and so
the above approximation scheme is inapplicable in this case.

\subsubsection{Conservation of angular momentum}\label{sec15}
\noindent From   (\ref{eqntid2snew}) 
we find that
\begin{align}
&\hspace{-5.9cm} \sum_{j=1}^{N} \frac{m_jM_j}{\lambda^2a_{j,0}} \frac{d\tilde z_j}{d\tau} =0.
\end{align}
This is a statement of the conservation of angular momentum in the small eccentricity limit
 as can be seen by writing it  in terms of unscaled variables in the form
\begin{align}
&\hspace{-5.9cm} \sum_{j=1}^{N} {m_j\sqrt{GM_j}}\left(\frac{d\sqrt{a_j}}{d t} \right)_0 = 0.
\end{align}
\section{Numerical simulations}\label{Numerics}
We now present simulations carried out adopting representations of the HD 158259 and EPIC 245950175 systems.
 {\bold{In these,  equations (\ref{emot}) - (\ref{Gammai}) were  solved }}as
 in previous work \citep[see eg.][]{Papaloizou2015, Papaloizou2016} {\bold{though in this case migration torques
 were not included.}}They were all initiated assuming zero eccentricities and random orbital phases.
 {\bold{ In particular we test the  predictions of the semi-analytic model described in Section 
 \ref{sec3}. }} 
Before describing the results for each system we give preliminary discussions of their main parameters.
\subsection{HD 158259}\label{sec16}
The parameters for this system are taken from \citet{Hara2020} and are listed 
in table \ref{table2}.

 \begin{table}[htbp]
 \centering
 \caption{Table \ref{table2}  Adopted planet parameters for the HD 158259 system}
\begin{tabular}{c|c} 
\hline
\hline
 \multicolumn{2}{c}{HD 158259} \\
\hline
Orbital period ({\it days}) & mass $(M_{\oplus})$   \\ 
\hline 
2.178 &  2.2  \\
\hline
3.432 &  5.6   \\
\hline
5.198 &  5.4  \\
\hline
7.951 &  6.1  \\
\hline
12.03 &  6.1  \\
\hline
17.42 &  6.9  \\
\hline
\end{tabular}\label{table2}
\end{table}

\noindent The period ratios  associated with 
consecutive pairs listed beginning with the innermost pair and moving outwards are
 $1.5758, 1.5146, 1.5296, 1.5130,$ and  $1.4480.$
 {\bold{In our simulations we  investigate secular evolution driven by dissipative tidal effects.
 As indicated by the semi-analytic model this is not  expected to depend on the initial orbital phases
 as was verified by considering a variety of simulations where these were are chosen at random  all of which yielded 
 qualitatively similar results.  The planets were not found to be in  mean motion resonance initially. We focus on representative cases below.}

Given the central mass $M=1.08M_{\odot}$ and adopting a characteristic planet mass
of $6 M_{\oplus}$ we  set  $\epsilon = 2.0\times10^{-5}.$
 {\bold {According to equation (\ref{scal4}) and the discussion immediately below that
 the choice of $\lambda$ should be such that
 $\epsilon^{2/3}/\lambda$ represents  an estimate of the fractional deviation from commensurability.
 Using equation (\ref{scal4}) with the above choice of $\epsilon,$ the above period ratios 
 indicate that
 $\epsilon^{2/3}/\lambda= 0.096, 0.019, 0.038, 0.017,$  and  $0.07.$
 The parameters $\lambda$ and $\epsilon$ were introduced as dimensionless parameters
 in an ordering scheme that should be small enough for the semi-analytic treatmment
 of Section \ref{sec2} to be applicable.
 The value of $\lambda$ should  indicate an order of magnitude and as it is not used in any calculation
 there is some latitude in its choice. 
 On this basis we make a representative choice for the single value
  $\lambda =0.02$ to define the scaling.  
      This small value is suggestive that the semi-analytic procedure discussed in Sections (\ref{sec12}) and (\ref{sec13})
for calculation of the resonant angle dynamics and epicyclic response is likely to be applicable.
This is explored by testing against the results of our simulations.}}
The evolution of the semi-major axes depends on  whether there are 
effective Laplace resonances  (see discussion in Section \ref{sec11}).
  \subsubsection{The possibility of  Laplace resonances}\label{sec17}
  For this system we find the three  three  planet relations that are closest to zero are
$(3n_3 -5n_2+2n_1)/2n_2=0.066,$
$(3n_4 -5n_3+2n_2)/2n_3=-4.8\times10^{-3}   $
and  $(3n_5 -5n_4+2n_3)/2n_4 = 0.02.$
 The vanishing of these would imply a strict Laplace resonance, 
These Laplace resonance conditions are satisfied with approximately the same precision
are  that of the first order resonances, the latter ranging between $0.  017$ and $0.096.$
Accordingly we might  expect the simple semi-analytic model to be applicable to the 
estimation of the rate of evolution of the semi-major axes.
\subsection{Simulation results}\label{sec18}
We present the result of  simulations with $Q'=1,$ and $Q'=2$ for all planets in the system.
An estimate of the mean density,  $\rho_{1} = 1.1\rho_{\oplus}$  is only available for the inner most planet.
In order to apply equation (\ref{teccsn}) for the circularisation time we assumed the same value for all planets.
Alternatively our specifications  can be regarded as equivalent to setting  $Q'(\rho_{j} /\rho_{\oplus})^{5/3}$ to be the same for each planet.
In that case the simulations can be regarded as being 
for $Q' = (1.1\rho_{\oplus}/\rho_{j})^{5/3}$ and $Q'= 2(1.1\rho_{\oplus}/\rho_{j})^{5/3}.$

\begin{figure}
\centering
\includegraphics[trim=0.0cm 0.00cm 0.00cm 0.00cm,clip=true, height=3.8in,angle=0]{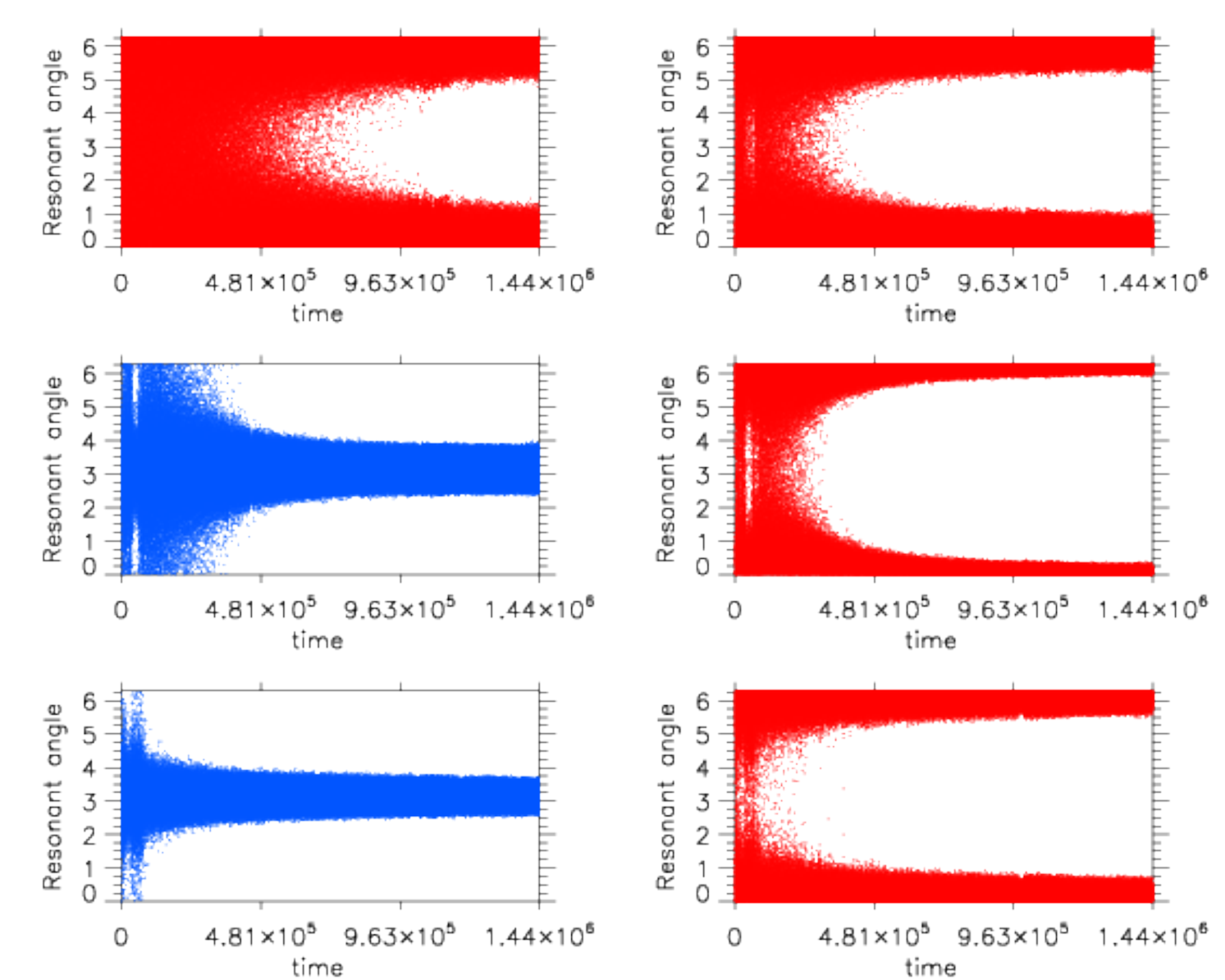}
\caption{\label{Fig1} The evolution of the resonant angles showing sustained libration for HD158259 with $Q'=1.$
In this figure and all those below times are expressed in years.
The top left panel shows $\Phi_{2,1,1}= 3\lambda_2 -2\lambda_1-\varpi_1.$ The top right panel shows $\Phi_{6,5,2}= 3\lambda_6 -2\lambda_5-\varpi_6.$
The leftmost  panel in  the middle row  shows $\Phi_{3,2,2}= 3\lambda_3 -2\lambda_2-\varpi_3.$  
The rightmost  panel in  the middle row  shows $\Phi_{3,2,1}= 3\lambda_3 -2\lambda_2-\varpi_2.$ 
The bottom left  panel   shows $\Phi_{5,4,2}= 3\lambda_5 -2\lambda_4-\varpi_5.$
 Finally the bottom right  panel   shows $\Phi_{5,4,1}= 3\lambda_5 -2\lambda_4-\varpi_4.$}
\end{figure}

\begin{table}[htbp]
 \centering
 \caption*{\textcolor{red}{Table \ref{table2r}   Librating resonant angles for the HD 158259 system}}
\textcolor{red}{\begin{tabular}{c|c|c|c} 
\hline
\hline
 \multicolumn{4}{c}{Resonant angles of the  form $\phi_{j+1,j,1}= 3\lambda_{j+1}-2\lambda_{j}-\varpi_{j} $} \\
\hline
Resonant   & Expected & $\left(\frac{\gamma_{j}/({\tilde\omega_{j+1,j})_0}}{\alpha_j/({\tilde\omega_{j,j-1}})_0}\right)^2$&
$sgn\left(\frac{\gamma_{j}}{({\tilde\omega_{j+1,j}})_0}\right)$  \\
 angle&libration center& &\\ 
\hline
$3\lambda_{2}-2\lambda_{1}-\varpi_{1}$ $(j=1)$ &  $0$ & $---$ & - \\
\hline
$3\lambda_{3}-2\lambda_{2}-\varpi_{2} $ $(j=2)$ &  $0$& $16.88$& -  \\
\hline
$ 3\lambda_{5}-2\lambda_{4}-\varpi_{4} $ $(j=4)$ & $0$& $6.34$ & - \\
\hline
\hline
\multicolumn{4}{c}{Resonant angles of the  form $\phi_{j,j-1,2}= 3\lambda_{j}-2\lambda_{j-1}-\varpi_{j} $} \\
\hline
Resonant   &Expected  & $\left(\frac{\alpha_{j}/({\tilde\omega_{j,j-1}})_0}{\gamma_j/({\tilde\omega_{j+1,j}})_0}\right)^2$&$sgn\left(\frac{\alpha_{j}}{({\tilde \omega_{j,j-1}})_0}\right)$  \\ 
angle&libration center&& \\
\hline 
 $3\lambda_{3}-2\lambda_{2}-\varpi_{3}$ $(j=3)$ & $ \pi$& $4.34$ & - \\
\hline
 $3\lambda_{5}-2\lambda_{4}-\varpi_{5}$ $(j=5)$ & $ \pi $ & $11.47$ &- \\
\hline
 $3\lambda_{6}-2\lambda_{5}-\varpi_{6}$ $(j=6)$ &  $0$  &$---$ &+\\
\hline
\hline
\end{tabular}}
 \caption{\textcolor{red}{
 \newline 
 Quantities associated with the determination of  whether resonant angles are expected to
  librate using the semi-analytic approach of Section \ref{sec13} in the limit of large circularisation times are tabulated.
  The first column gives the resonant angle. This is of either the form 
   $\phi_{j+1,j,1}= 3\lambda_{j+1}-2\lambda_{j}-\varpi_{j}$ 
   or $\phi_{j,j-1,2}= 3\lambda_{j}-2\lambda_{j-1}-\varpi_{j}.$
   The second column gives the derived libration center.
  The third column gives $\left((\gamma_{j}/({\tilde\omega_{j+1,j}})_0)/(\alpha_j/({\tilde\omega_{j,j-1}})_0)\right)^2$ in the case of angles
  of the form $\phi_{j+1,j,1}$ and $\left((\alpha_{j}/({\tilde \omega_{j,j-1}})_0)/(\gamma_j/({\tilde \omega_{j+1,j}})_0)\right)^2$ in the case of angles of the
  form  $\phi_{j,j-1,2}.$  The angles for which these quantities can be defined are expected to librate when they exceed unity but
  they play no role when the resonance involves the innermost or outermost planet (see 
  discussion in  Section \ref{sec13}  and in particular 
  equations (\ref{condlib1}) and
  (\ref{condlib2})).  The fourth column gives the  either the sign of  $\gamma_{j}/({\tilde \omega_{j+1,j}})_0$  or     $\alpha_{j}/({\tilde\omega_{j,j-1}})_0$.
  which determine the center of libration for the associated angles as described in Section \ref{sec13}. 
  Only the angles that librate are considered. These are shown in Fig.\ref{Fig1}.}}\label{table2r}
\end{table}

\begin{figure}
\centering
\includegraphics[trim=0.0cm 0.00cm 0.0cm 0.0cm,clip=true, height=3.8in,angle=0]{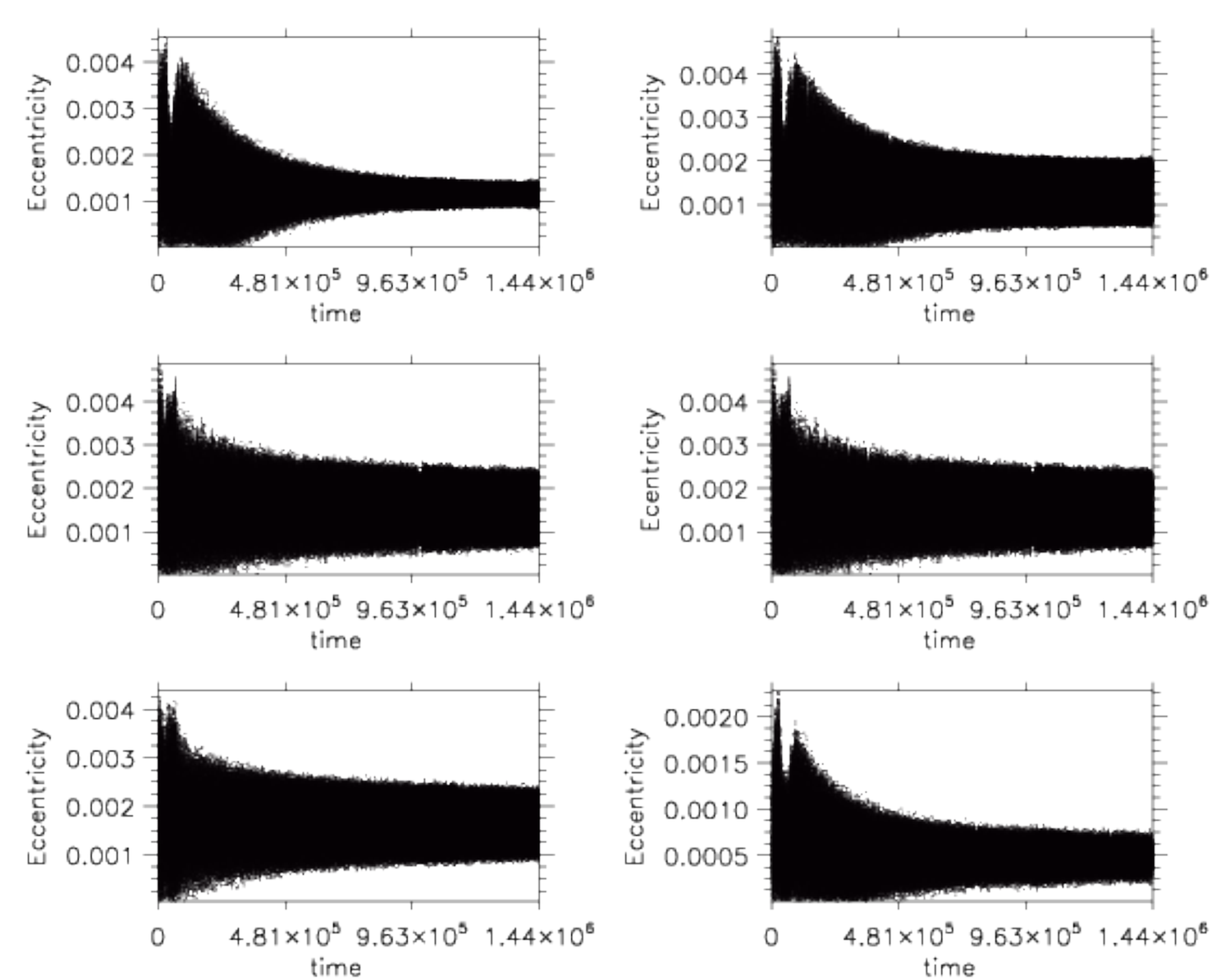}
\caption{\label{Fig2}The evolution of the eccentricities  for HD 158259 with $Q'=1.$
The top left, top right, middle left, middle right, bottom left and bottom right panels respectively
show the eccentricities of the six planets  starting from the innermost and moving consecutively to the outermost.} 
\end{figure}

\begin{figure}
\centering
\includegraphics[trim=0.0cm 0.00cm 0.0cm 0.0cm,clip=true, height=1.6in,angle=0]{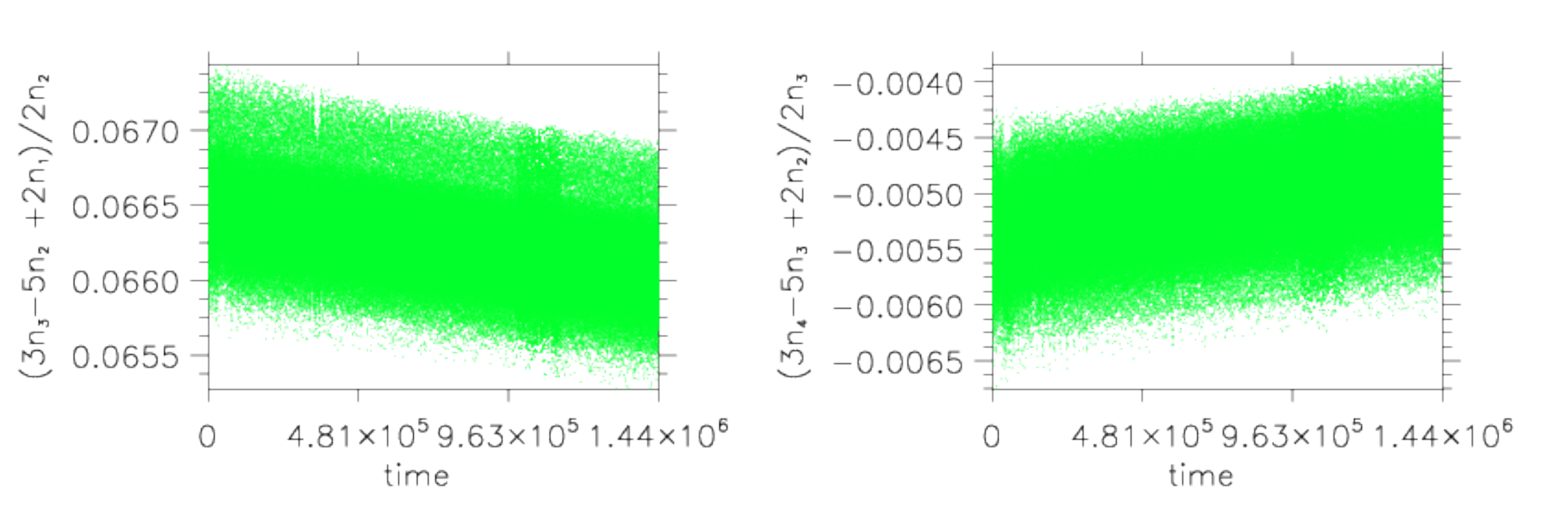}
\caption{\label{Fig3} The evolution of  $(3n_3-5n_2+2n_1)/(2n_2)$ (left panel) and\newline
$(3n_4-5n_3+2n_2)/(2n_3)$ (right panel) for HD 158259 and $Q'=1.$
These quantities would vanish in the limit of small eccentricities if there was a strict Laplace resonance between 
the innermost three planets in the former case and the second, third and fourth 
innermost  planets in the latter case. In this case  fluctuations in these quantities are relatively small compared to their deviations
from zero as they evolve.
}
\end{figure}

\begin{figure}
\centering
\includegraphics[trim=3.0cm 0.00cm 0.0cm 0.0cm,clip=true, height=4.0in,angle=0]{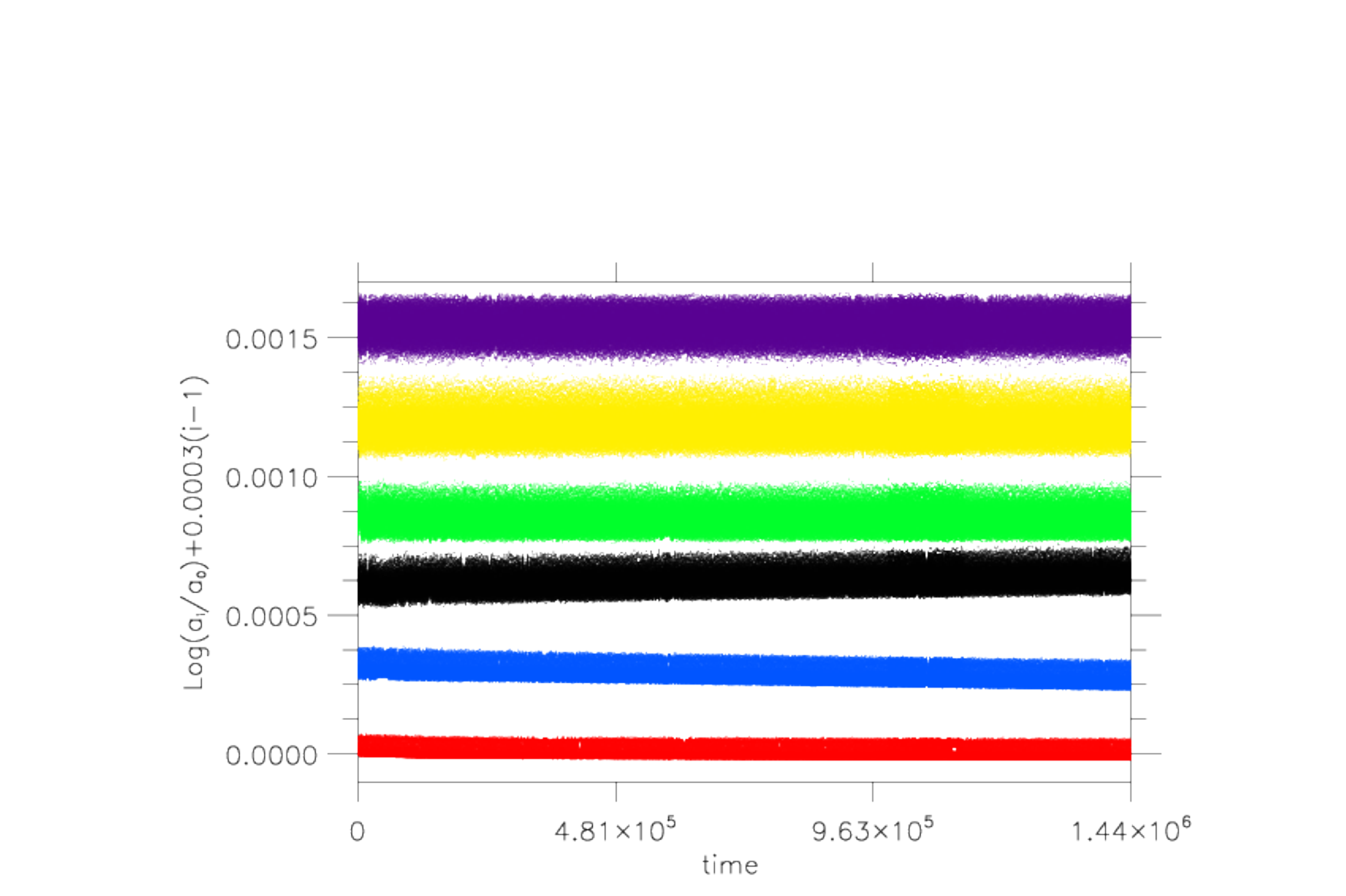}
\caption{\label{Fig4} The evolution of the semi-major axes for HD 158259 and $Q'=1.$
the quantities, $\log(a_i/a_0)+0.0003(i-1),$ where $a_i$ is the semi-major axis of planet $i, i = 1,2...6,$
and $a_0$ refers to  its initial value.  The plots are for planets, $i=1$ to $i=6,$ moving consecutively from the lowermost (red)
to the uppermost (majenta).}
\end{figure}

\begin{figure}
\centering
\includegraphics[trim=0.0cm 0.00cm 0.00cm 0.00cm,clip=true, height=3.8in,angle=0]{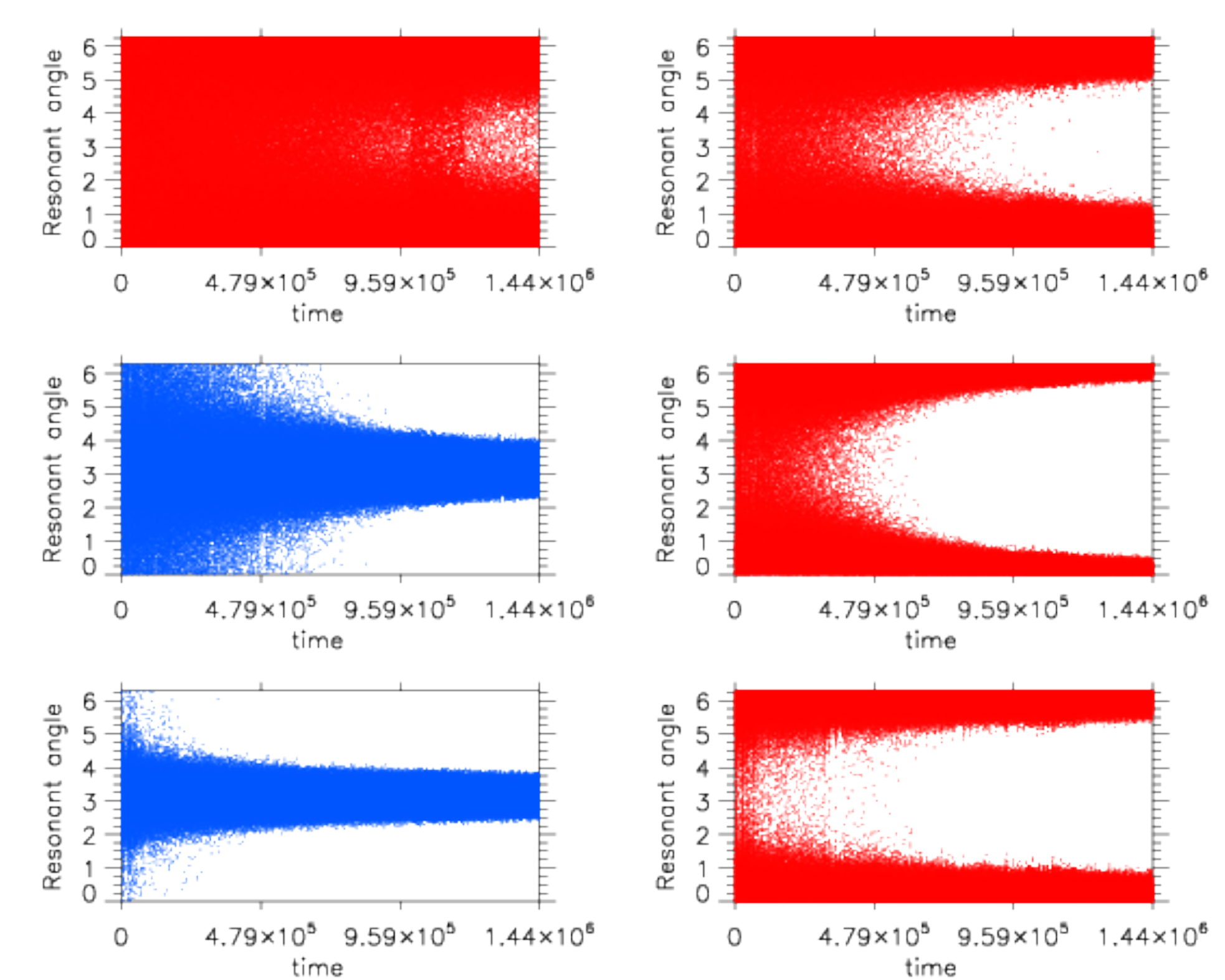}
\caption{\label{Fig5}As in Fig.\ref{Fig1} but for $Q'=2.$}
\end{figure}

\begin{figure}
\centering
\includegraphics[trim=0.0cm 0.00cm 0.0cm 0.0cm,clip=true, height=3.8in,angle=0]{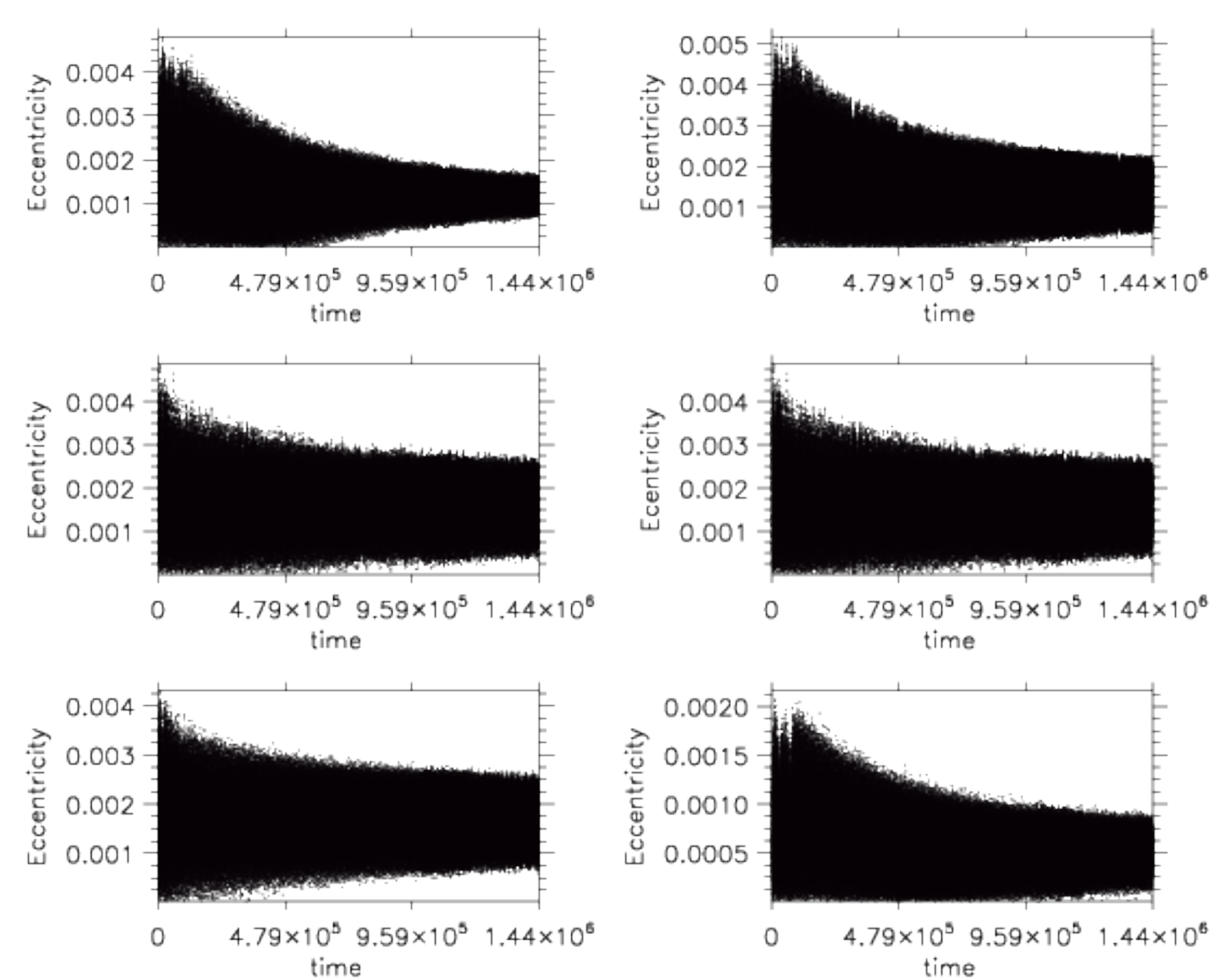}
\caption{\label{Fig6} As in \ref{Fig2} but for $Q'=2.$}
\end{figure}

\begin{figure}
\centering
\includegraphics[trim=0.0cm 0.00cm 0.0cm 0.0cm,clip=true, height=1.6in,angle=0]{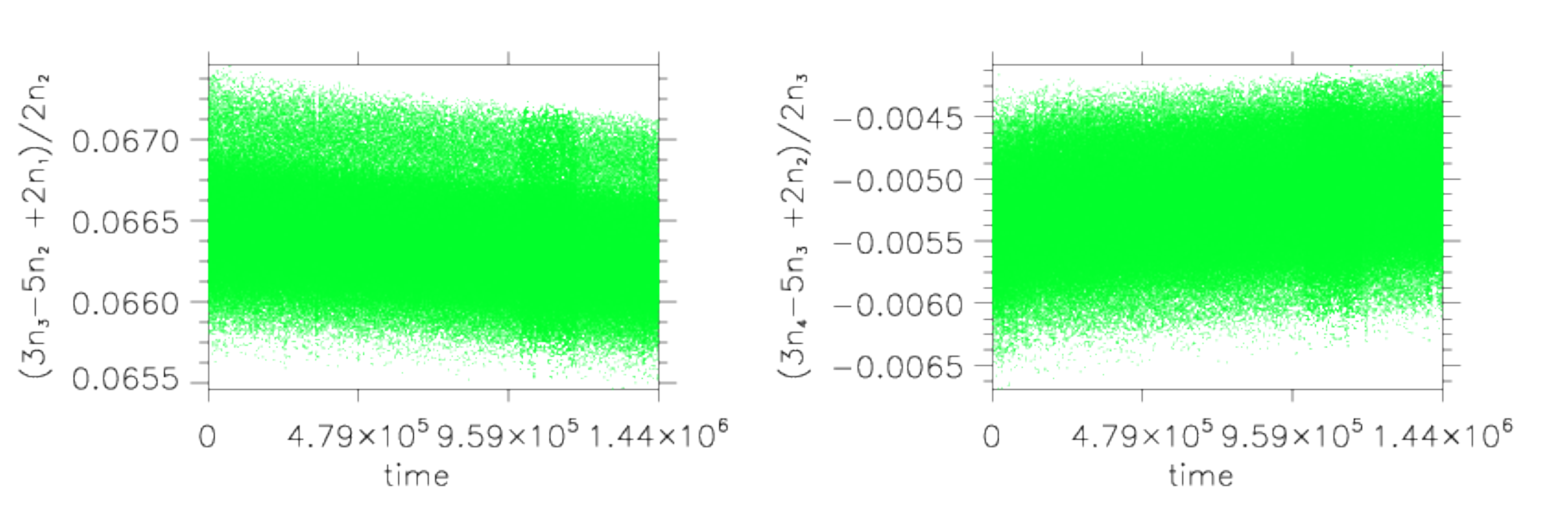}
\caption{\label{Fig7}As in Fig. \ref{Fig3} but for $Q'=2.$}
\end{figure}

\begin{figure}
\centering
\includegraphics[trim=3.0cm 0.00cm 0.0cm 0.0cm,clip=true, height=4.0in,angle=0]{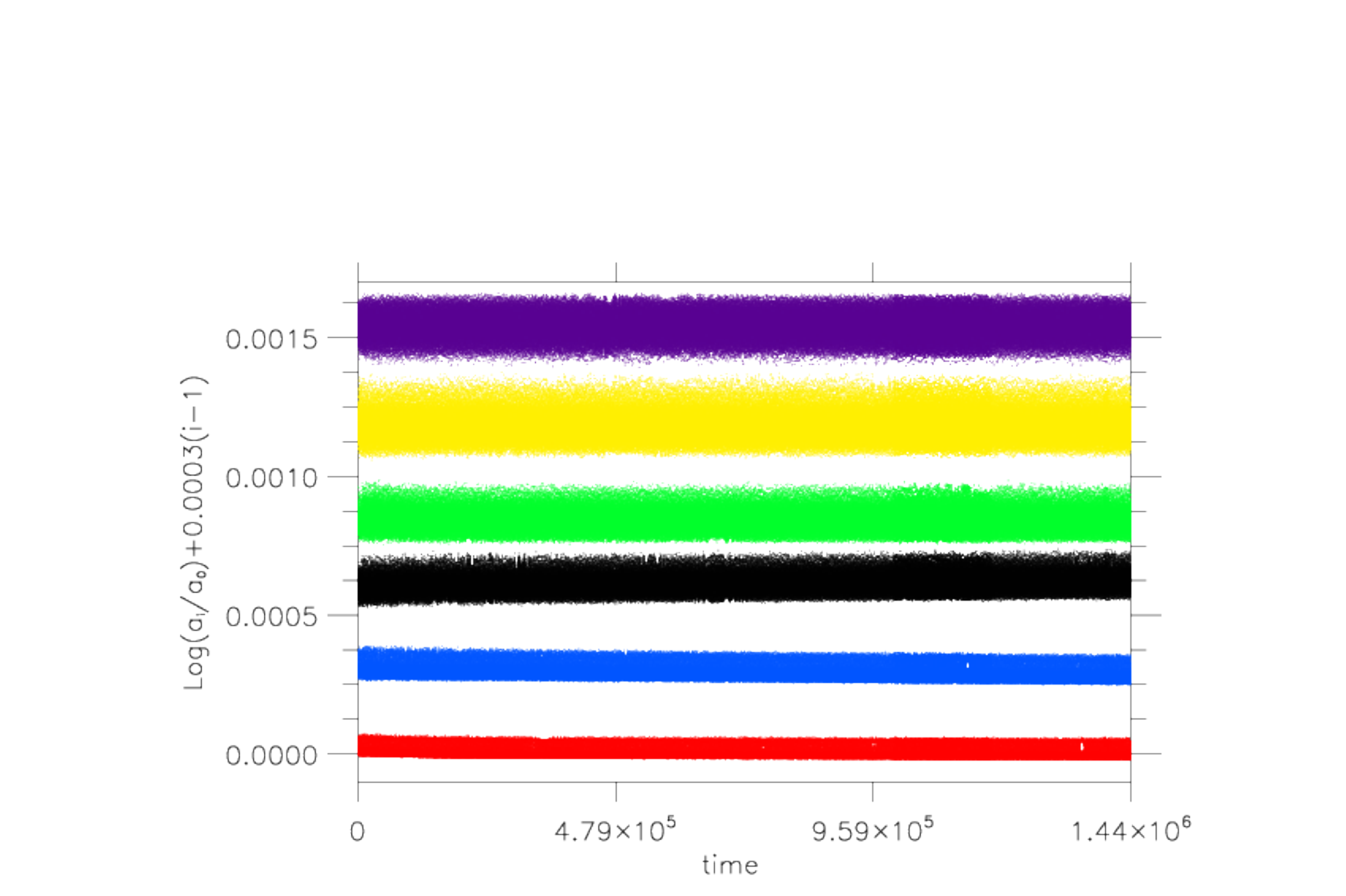}
\caption{\label{Fig8} As in Fig. \ref{Fig4} but for $Q'=2.$}
\end{figure}

In Fig. \ref{Fig1} we show the evolution of the resonant angles that end in clear libration
after $\sim 1.4\times 10^6 y$ for the case with $Q'=1$ having started with random orbital phases. These are
$\Phi_{2,1,1}= 3\lambda_2 -2\lambda_1-\varpi_1.$  $\Phi_{6,5,2}= 3\lambda_6 -2\lambda_5-\varpi_6.$
 $\Phi_{3,2,2}= 3\lambda_3 -2\lambda_2-\varpi_3,$  
 $\Phi_{3,2,1}= 3\lambda_3 -2\lambda_2-\varpi_2.$ 
 $\Phi_{5,4,2}= 3\lambda_5 -2\lambda_4-\varpi_5.$
 and $\Phi_{5,4,1}= 3\lambda_5 -2\lambda_4-\varpi_4.$
 Note that there are short period fluctuations in these quantities  
in this and other figures that are not resolved on the scale shown.
Notably regular oscillations are expected from the forced eccentricities 
determined in Section \ref{sec12}.
In addition to these there are other fluctuations neglected in the averaging process that
led to the simplified  model equations. These may be  crudely characterised by considering the
parameter $ f_{sc} = 2G\epsilon M /(\Delta v_R^2),$ with $v_R$ and $\Delta$ being the relative velocity and distance
of closest approach of neighbouring planets, 
{\bold{here assumed to be initially on orbits that can be assumed to be circular.
When this dimensionless quantity $\ll 1$ it  measures twice the magnitude of the fractional change in the relative velocity
that occurs were the gravitational interaction
between the planets during closest approach treated as a simple two body scattering  with the central mass and other planets
being neglected \citep[see eg.][]{Papaloizou1979}. 
Note that  this change is induced  during the  phase of the encounter prior to closest approach
and then it is subsequently reversed. Net changes of the semi-major axes as a result of the encounter are  found to be second order in $f_{sc}$  \citep[see][]{Papaloizou1979}.
For planet $j,$ $f_{sc},$  }}  may also be written as
$f_{sc} = 8\epsilon(a_j/\Delta)^{3}/9.$  Adopting $\epsilon =2\times 10^{-5}$ and $\Delta/a_j  =0.25,$
$f_{sc}$ is estimated to be $\sim  0.0011.$ {\bold{The magnitude of the expected  relative excursion the semi-major axis
is $\sim f_{sc} |v_R| \sqrt{a_j/GM} \sim 1.5f_{sc}\Delta/a_j \sim 0.0004$
The characteristic relative excursions of the semi-major axes of
 the six planets  in the system illustrated  in  the discussion of the evolution of the semi-major axes presented below 
 are found to  vary between $\sim 0.0002$
and  $0.0006.$   Thus there
appears to be  consistency with the simulations given the approximations made 
in order to obtain the estimates  in the above discussion.}}

{\bold{ We used  results of the semi-analytic theory discussed in Section \ref{sec13} to determine which  of the resonant angles 
$\Phi_{j+1,j,1}  ,1\le j\le 5,$ or $\Phi_{j,j-1,2}  ,2\le j\le 6,$ were expected to librate.
The criteria adopted  are given by equation (\ref{condlib1}) in the former case and equation (\ref{condlib2}) in the
latter case. Whether the libration is about zero  or $\pi$
in the former case was specified   according to whether $ \gamma_j/({\tilde \omega_{j+1,j} })_0$ is negative or positive,
and in the latter case according to whether $ \alpha_j/(\tilde \omega_{j,j-1})_0$ was positive or negative.
\textcolor{red}{Some of the parameters involved are tabulated in table \ref{table2r}.}
 Note too that the above criteria do not depend on the values of the scaling parameters $\epsilon$ and $\lambda$
as these cancel out. We remark that they correctly predict the libration of  $\Phi_{6,5,2}$  associated with planets 5 and 6
even though the departure of these from commensurability is significantly greater than for other pairs.
In this context we remark that  libration may still occur for such moderately large departures \citep[see eg.][]{Papaloizou2011, Papaloizou2015}.
 }} 
Our numerical results were
found to be fully  consistent with the {\bold{ above \newline determinations and the}} discussion in Section \ref{sec13} {\bold{thus confirming}} the applicability
of  the simple analytic model.

The evolution of the eccentricities  for the six planets is illustrated in Fig.\ref{Fig2}.
Their characteristic values are steady and  $\sim 0.001.$ However, fluctuations  can reduce them
to near zero. Root mean square eccentricities for  planets  $(j = 1 - 6) $  estimated from 
the analysis in Section \ref{sec12}  are respectively  $0.0002, 0.0017, 0.00148, 0.00163, 0.00167$  and $0.00045.$
Corresponding measurements of $0.7\times$ steady maximum  values are
$0.0011, 0.0014, 0.0018,0.0018,0.0018$ and $0.00053.$ which are similar  but with the largest
discrepancy applying to the innermost planet. This is likely to be because this planet is the furthest from resonance
making the estimated eccentricity smaller in magnitude in comparison to {\bold{that induced by neglected effects.}}

The evolution of  the quantities $(3n_3-5n_2+2n_1)/(2n_2)$ and  \newline
$(3n_4-5n_3+~2n_2)/(2n_3)$   are  illustrated in Fig. \ref{Fig3}.
It can be seen that although there are fluctuations in these quantities, their amplitude is
relatively small compared to the distances of their means from zero.
It can also be seen that the means are slowly evolving towards zero
which will be attained {\bold{more quickly in the latter case}}
on a characteristic time scale $\sim 2\times 10^7 Q'y,$
where we have assumed scaling {{\boldmath of this evolution time scale}}. with $Q'$ (see below).
If the system was formed with orbital periods close to their present values,
in order to avoid being significantly closer to strict Laplace resonances,
the above discussion indicates that we require $ Q' > 100 t_a/(2\times10^{9} y)$
where $t_a$ is {\bold{ the time since formation}}. 

The evolution of the semi-major axes for the six planets
is illustrated in  Fig.~\ref{Fig4} from which it can be seen that, 
after averaging out fluctuations, the innermost two
are moving inwards and the next planet  $(j=3)$  is moving outwards.
Any secular  movement of the outer planets is significantly smaller.

For the innermost planet $(j=1)$ the value
 $d\log(a)/dt~=~-~5.4~\times~10^{-12}~y^{-1}$ 
was measured from the simulation, whereas the value 
 estimated  from (\ref{eqntid2snew}) after removing the variable scaling (and hence the parameters $\epsilon$ and $\lambda$) was
$d\log(a)/dt=-3.4 \times 10^{-12} y^{-1}.$

The corresponding values for planet $(j=2)$ were respectively 

\noindent $d\log(a)/dt= -2.8\times10^{-11} y^{-1}$
and $d\log(a)/dt= -2.8\times10^{-11} y^{-1}.$
For planet $(j=3)$ they were respectively $d\log(a)/dt= 2.8\times10^{-11} y^{-1}$
and 

\noindent $d\log(a)/dt= 2.5\times10^{-11} y^{-1}.$

For planets $ (j=4,5,$ and $6)$  values could not be reliably measured while 
  very small values  were estimated from  (\ref{eqntid2snew}), being respectively,
$d\log(a)/dt= -3.1\times10^{-13} y^{-1}, $ $d\log(a)/dt= 1.1\times10^{-12} y^{-1},$ and 

\noindent $d\log(a)/dt= 1.8\times10^{-14} y^{-1}.$

These results indicate that the dominant evolution will be the inward migration of the two inner most planets
balanced by the outward migration of the third planet $(j=3).$ The rates of evolution determined from the
simulation and the simple semi-analytic model are in reasonable agreement 
with the innermost planet's migration being somewhat underestimated in the latter case. This may be on account
of the distance of this planet from commensurability as indicated above. The most rapid inward migration occurred for the second innermost planet
being on a time scale $\sim 1.6\times 10^{10}Q'  y.$ 

In order to check the scaling of the above results with $Q',$ we have repeated the above simulation with $Q'=2$
and the results corresponding to Figs. \ref{Fig1} - \ref{Fig4} are illustrated in Figs. \ref{Fig5} - \ref{Fig8}.
As expected the evolution of the semi-major axes is consistent with being slowed down by a factor of two
as is the evolution of the resonant angles and  eccentricities. In particular the resonant angle\newline 
 $\Phi_{2,1,1}= 3\lambda_2 -2\lambda_1-\varpi_1$
only starts to enter libration at the end of the simulation while the eccentricities eventually attain  similar values but more slowly. 

\subsection{ EPIC 245950175}\label{sec19}
The parameters for this system {\bold {also known as K2-138}} are taken from \citet{Lopez2019} and are listed 
in table \ref{table3}.
 \begin{table}[htbp] 
\centering
\caption{Table \ref{table3} Adopted planet parameters for the EPIC 245950175 system}\label{table3}
\begin{tabular}{c|c|c} 
\hline
\hline
     \multicolumn{3}{c}{EPIC 245950175} \\
\hline
Orbital period ({\it days}) & mass $(M_{\oplus})$  &  mean density $(\rho_{\oplus})$  \\ 
\hline 
2.353&  3.1  & 0.88   \\
\hline
3.560 &  6.3  & 0.51   \\
\hline
5.405&  7.9  & 0.57  \\
\hline
8.261 &  13.0  & 0.33 \\
\hline
12.76&  1.6  & 0.064   \\
\hline
41.97 &  4.3  & 0.16  \\
\hline
\end{tabular}
\end{table}
The period ratios  associated with 
consecutive pairs listed beginning with the innermost pair and moving outwards 
are $1.5129, 1.5183, 1.5284, 1.5446$ and  $3.289.$
{\bold{The same considerations apply to these simulations as to those of HD 158259.
As for that system the planets were found not to be in mean motion resonance initially.}}

Given the central mass $M=0.98M_{\odot}$ and adopting a characteristic planet mass
of $6 M_{\oplus}$ we set  $\epsilon = 2.0\times10^{-5}.$
Using the relation (\ref{scal4}), the above period ratios suggest
 $\epsilon^{2/3}/\lambda= 0.017, 0.024, 0.037,$ and  $0.058$ 
  as being appropriate to the four 
  consecutive pairs starting with the innermost pair and moving outwards
 and thus we make the  representative choice for the single value
  $\lambda =0.03$ to define the scaling.  
   We do not consider the outermost pair in the above discussion as they are not near a first order resonance
   and thus the outermost planet is found  not to  contribute significantly to the dynamics of the inner ones.
   This discussion indicates that the simple procedure discussed in Sections \ref{sec12} and \ref{sec13}
for the  calculation of the epicyclic  and resonant angle dynamics should be applicable.
However, this is not the case for the  evolution of the semi-major axes on account of the
effect of  Laplace resonances  (see discussion in Section \ref{sec11}).
  \subsubsection{Potential Laplace resonances}\label{sec20}
  For this system we find the three planet relations, the vanishing of which imply a strict Laplace resonance, 
$(3n_3 -5n_2+2n_1)/2n_2=8.47\times10^{-4},$\newline
$(3n_4 -5n_3+2n_2)/2n_3=-2.82\times10^{-4}, $
and  $(3n_5 -5n_4+2n_3)/2n_4 =-4.74~\times~10^{-4}.$
In contrast to the HD 158259 system,
the Laplace resonance conditions are satisfied with a significantly greater precision than 
are the first order resonances. Typically the ratio of the deviations is $\sim 10^{-2}$
and they exceed $\lambda^3$ by around only one order of magnitude.
In addition the magnitude of these  deviations turns out to be less than that associated with short
term variations in the semi-major axes (see below).

\begin{table}[htbp]
 \centering
 \caption*{\textcolor{red}{Table \ref{table3r}  Librating resonant angles for the EPIC 245950175 system }}
\textcolor{red}{\begin{tabular}{c|c|c|c} 
\hline
\hline
 \multicolumn{4}{c}{Resonant angles of the  form $\phi_{j+1,j,1}= 3\lambda_{j+1}-2\lambda_{j}-\varpi_{j} $} \\
\hline
Resonant   &Expected  & $\left(\frac{\gamma_{j}/({\tilde\omega_{j+1,j}})_0}{\alpha_j/({\tilde\omega_{j,j-1}})_0}\right)^2$&
$sgn\left(\frac{\gamma_{j}}{({\tilde \omega_{j+1,j}})_0}\right)$  \\ 
angle &libration center&&\\ 
\hline
$3\lambda_{2}-2\lambda_{1}-\varpi_{1}$ $(j=1)$ &  $0$ & $---$ & - \\
\hline
$3\lambda_{3}-2\lambda_{2}-\varpi_{2}$ $(j=2)$ &  $0$& $2.788$& -  \\
\hline
$ 3\lambda_{4}-2\lambda_{3}-\varpi_{3}$ $(j=3)$ & $0$& $1.44$ & - \\
\hline
\hline
 \multicolumn{4}{c}{Resonant angles of the  form $\phi_{j,j-1,2}= 3\lambda_{j}-2\lambda_{j-1}-\varpi_{j} $} \\
\hline
Resonant  &Expected  & $\vspace{0mm}\left(\frac{\alpha_{j}/({\tilde \omega_{j,j-1}})_0}{\gamma_j/({\tilde\omega_{j+1,j}})_0}\right)^2$&$sgn\left(\frac{\alpha_{j}}{({\tilde \omega_{j,j-1}})_0}\right)$  \\ 
angle & libration center & &  \\
\hline 
 $3\lambda_{4}-2\lambda_{3}-\varpi_{4}$ $(j=4)$  & $ \pi$& $75.67$ & - \\
\hline
 $3\lambda_{5}-2\lambda_{4}-\varpi_{5}$ $(j=5)$ & $ \pi $ & $3.985\times10^5$ &- \\
\hline
\hline
\end{tabular}}
\caption{\textcolor{red}{
\newline
 Quantities associated with the determination of  whether resonant angles are expected to
  librate using the semi-analytic approach of Section \ref{sec13} in the limit of large circularisation times are tabulated
  as in table \ref{table2r} but for the EPIC 245950175 system. The resonant angles considered are shown in Fig. \ref{Fig9}.} \label{table3r}}
\end{table}

\begin{figure}
\centering
\includegraphics[trim=0.0cm 0.00cm 0.00cm 0.00cm,clip=true, height=3.8in,angle=0]{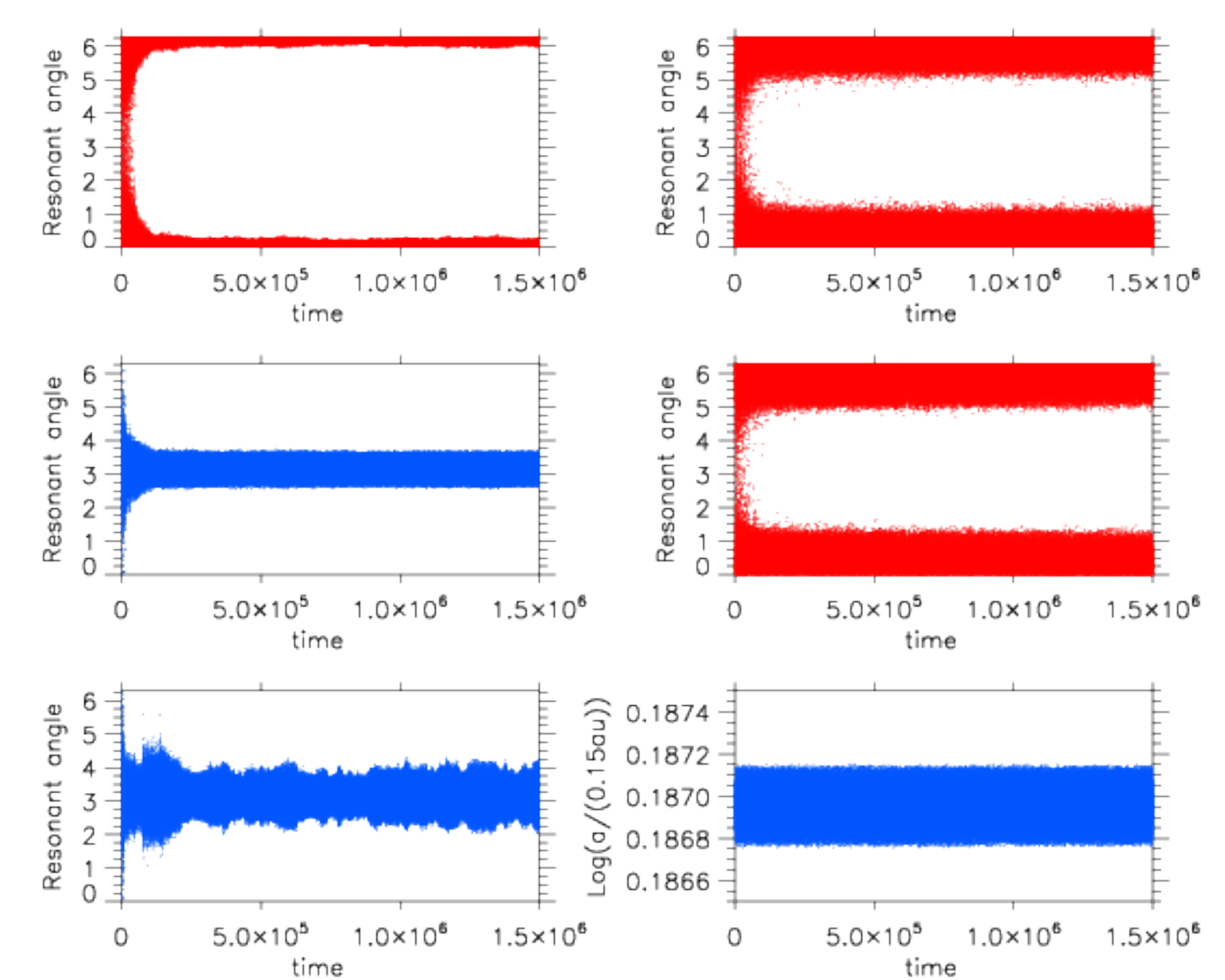}
\caption{\label{Fig9}The evolution of the resonant angles showing sustained libration for EPIC 245950175 with $Q'=1.$
The top left panel shows $\Phi_{2,1,1}= 3\lambda_2 -2\lambda_1-\varpi_1.$ The top right panel shows $\Phi_{3,2,1}= 3\lambda_3 -2\lambda_2-\varpi_2.$
The leftmost  panel in  the middle row  shows $\Phi_{4,3,2}= 3\lambda_4 -2\lambda_3-\varpi_4.$  
The rightmost  panel in  the middle row  shows $\Phi_{4,3,1}= 3\lambda_4 -2\lambda_3-\varpi_3.$ 
The bottom left  panel   shows $\Phi_{5,4,2}= 3\lambda_5 -2\lambda_4-\varpi_5.$ 
Finally the bottom right panel shows the evolution of $\log(a/(0.15au),$ $a$ being the semi-major axis for the outermost planet.
This undergoes small amplitude oscillations  with negligible mean evolution and does not significantly affect
the evolution of the inner planets.}
\end{figure}

\begin{figure}
\centering
\includegraphics[trim=0.0cm 0.00cm 0.0cm 0.0cm,clip=true, height=3.8in,angle=0]{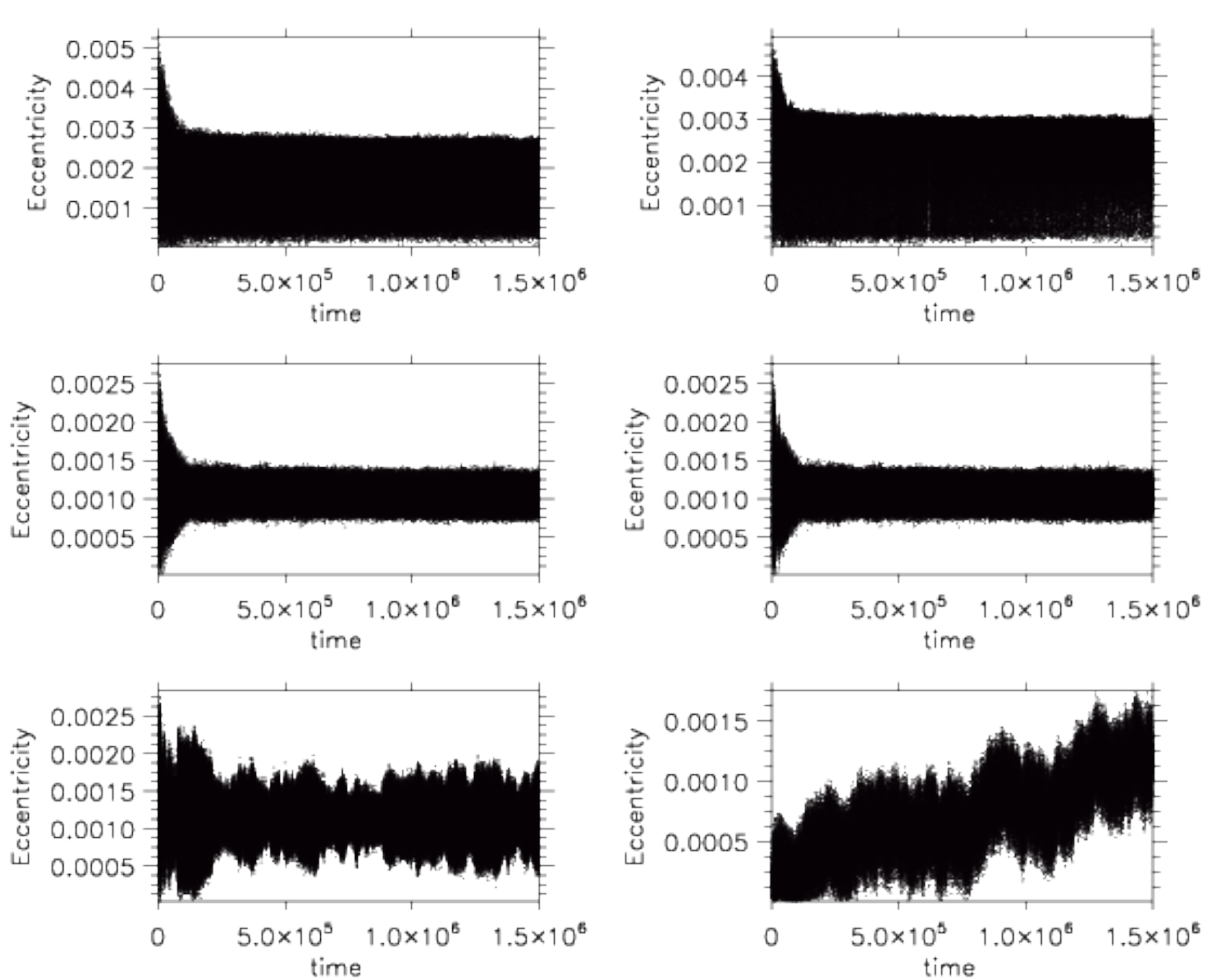}
\caption{\label{Fig10} As in Fig. \ref{Fig2} but  for EPIC 245950175 with $Q'=1.$}
\end{figure}

\begin{figure}*
\centering
\hspace{-0.0cm}
\includegraphics[trim=.3cm 0.00cm 0.0cm 0.0cm,clip=true, height=6.3in, width=4.7in, angle=0]
{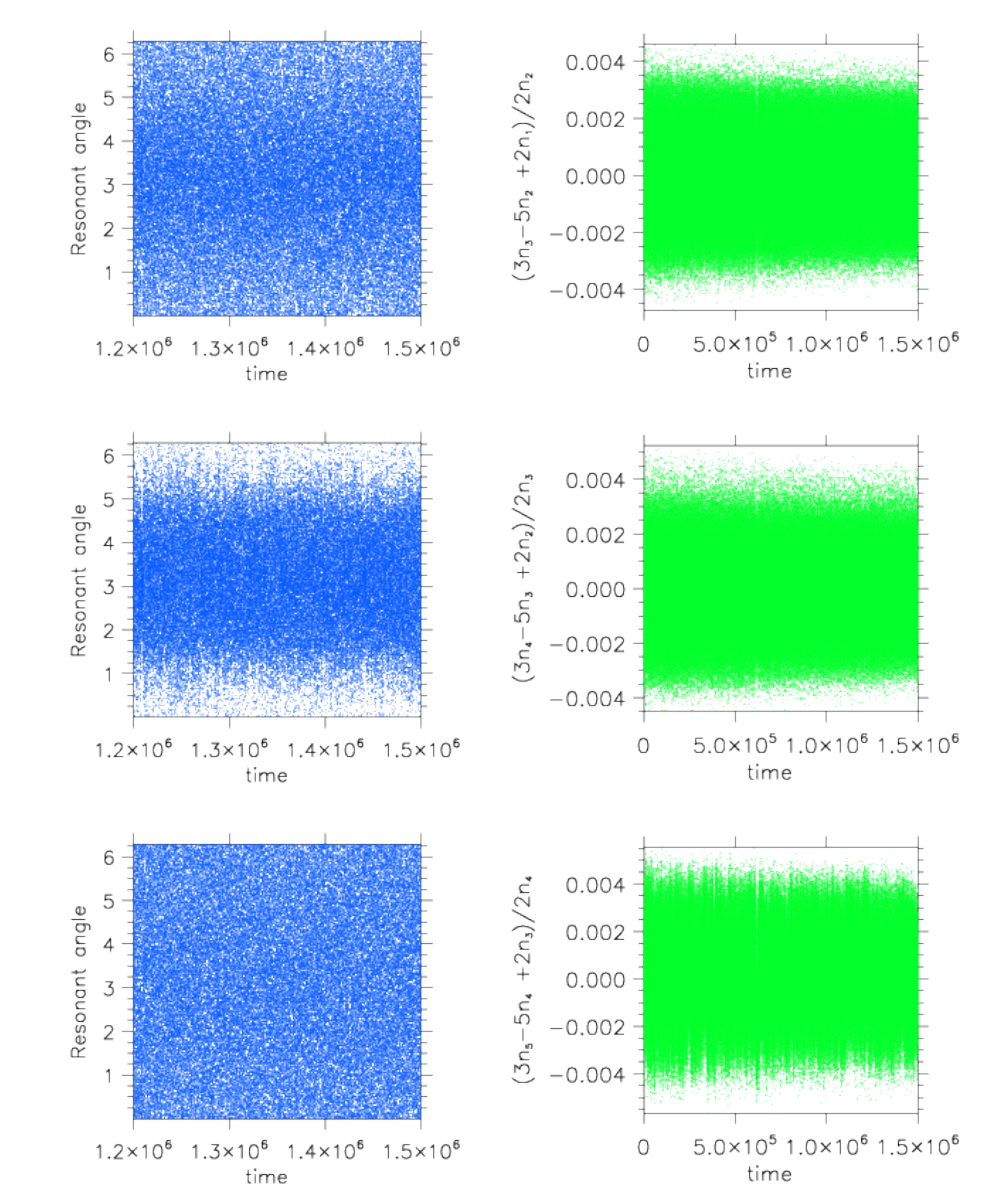}
\caption{\label{Fig11}.
The evolution of the resonant angles 
 {\bold{  
 $\Phi_{3,2,1} -\Phi_{2,1,2} = 3\lambda_3 -5\lambda_2 +2\lambda_1$ (top left panel), 
  $\Phi_{4,3,1} -\Phi_{3,2,2} = 3\lambda_4 -5\lambda_3 +2\lambda_2$ (left middle  panel ) and   
  $\Phi_{5,4,1} -\Phi_{4,3,2} = 3\lambda_5 -5\lambda_4 +2\lambda_3$ (bottom left panel) 
  towards the end of the simulation
 for EPIC 245950175 
with $Q'=1.$ Although these do not  librate the distribution of the first two over $(0,2\pi) $
exhibit  a  slight degree of nonuniformity (see text for more detail).}}
The upper right, middle right,  and lower  right panels 
respectively show the evolution of   $(3n_3-5n_2+2n_1)/(2n_2),$ 
$(3n_4-5n_3+2n_2)/(2n_3)$  and  $(3n_5-5n_4+2n_3)/(2n_4).$ 
In this case these three quantities maintain a mean value {\bold{ that stays in the neighbourhood of}} 
zero throughout the evolution indicating the significance
of the corresponding  Laplace resonances. }
\end{figure}

\begin{figure}
\centering
\includegraphics[trim=3.0cm 0.00cm 0.0cm 0.0cm,clip=true, height=4.0in,angle=0]{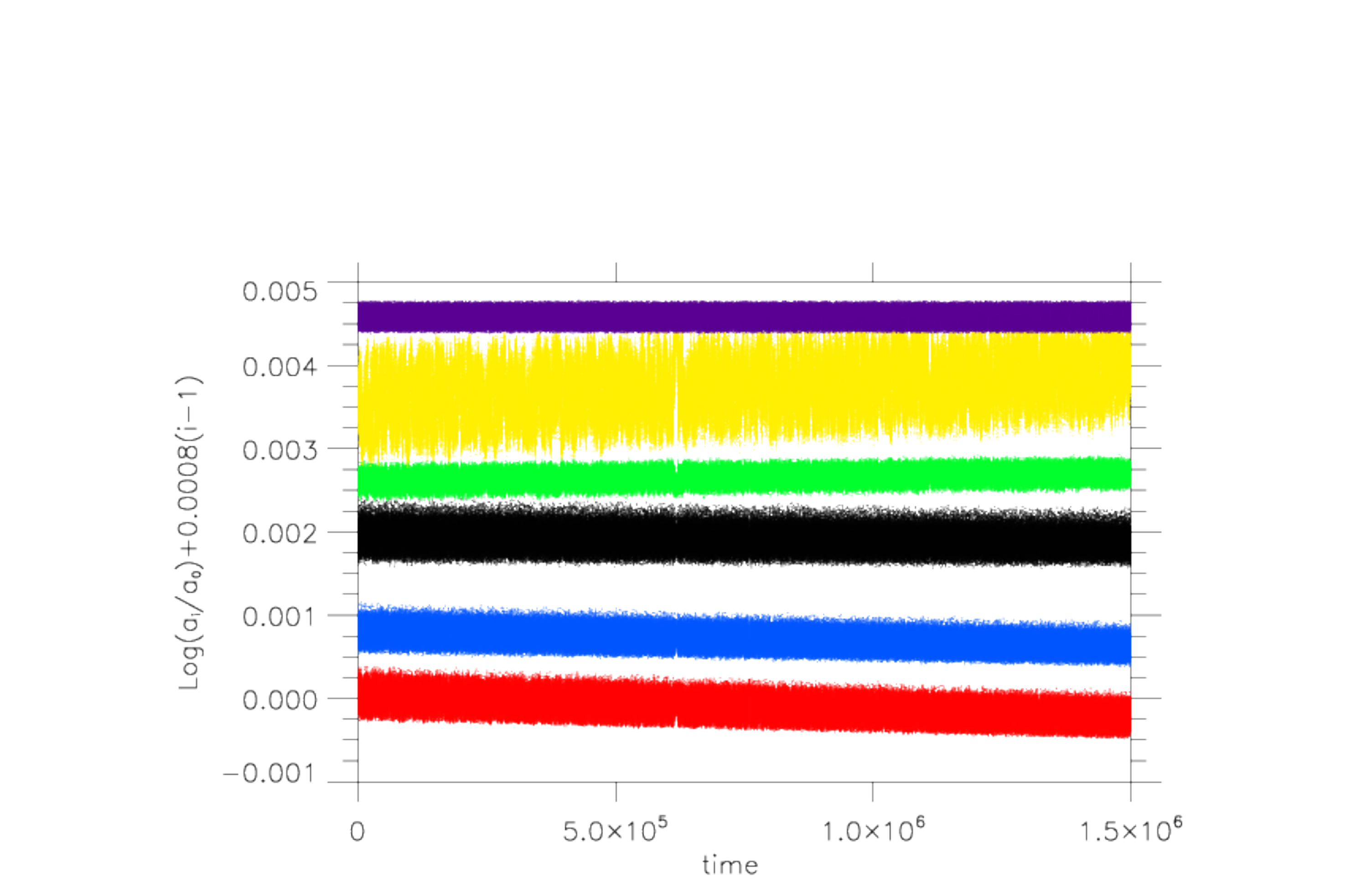}
\caption{\label{Fig12}  As in Fig. \ref{Fig4} but  for EPIC 245950175 with $Q'=1.$}
\end{figure}

\subsection{Simulation results}\label{sec21}

We have performed sets of simulations with $Q'=1,$ and $Q'=3.$
In Fig. \ref{Fig9} we show the evolution of the resonant angles that end in clear libration
after $\sim 1.5\times 10^6 y$  for the case with $Q'=1.$ These are
 $\Phi_{2,1,1}= 3\lambda_2 -2\lambda_1-\varpi_1,$ $\Phi_{3,2,1}= 3\lambda_3 -2\lambda_2-\varpi_2,$
 $\Phi_{4,3,2}= 3\lambda_4 -2\lambda_3-\varpi_4,$  
$\Phi_{4,3,1}= 3\lambda_4 -2\lambda_3-\varpi_3$. and 
$\Phi_{5,4,2}= 3\lambda_5 -2\lambda_4-\varpi_5.$ 
 The expected  libration of the above angles and whether the libration is about zero is 
again found to be fully consistent with the discussion in Section \ref{sec13} and confirms the applicability
of  the simple analytic model on this context.
\textcolor{red}{ Some of the parameters involved are tabulated in table \ref{table3r}. }

The evolution of $\log(a/(0.15au),$ $a$ being the semi-major axis for the outermost planet is also shown.
This planet is non resonant and plays only a small role in the evolution of the inner planets.
Its semi-major axis shows negligible change in the mean. 
 

The evolution of the eccentricities  for the six planets is illustrated in Fig.\ref{Fig10}.
Their characteristic values are steady and  $\sim 0.001.$ However, fluctuations  can reduce them
to near zero in some cases. Root mean square eccentricities for  planets  $(j = 1 - 5) $  estimated from 
the analysis in Section \ref{sec12}  are respectively  $0.0018, 0.0018, 0.0021, 0.0011$ and  $0.0011.$  
Corresponding measurements of $0.7\times$ steady maximum  values, also being approximate mean values
in the cases of planets  $(j=3)$ and $(j=4),$  are respectively
$0.0020, 0.0020, 0.0010,0.0010 $ and $0.0013.$ which are similar  but with the largest
discrepancy applying to the third innermost planet. This is likely to be associated with the effects of a Laplace resonance
producing a significant effect on its migration (see below).
The outermost planet attains eccentricities up to $0.001$ but these are through non resonant interactions.

The evolution of  the quantities
 $(3n_3-5n_2+2n_1)/(2n_2),$ \newline
  $(3n_4-5n_3+2n_2)/(2n_3)$  and  $(3n_5-5n_4+2n_3)/(2n_4).$ 
are shown in Fig. \ref{Fig11}. The vanishing of these indicates the presence of a Laplace resonance.
Here these three quantities maintain a mean value {\bold{that remains close to}} zero throughout the evolution indicating the significance
of these resonances. This is in contrast to the HD 158259 system  discussed above even though the migration rates
are comparable or faster in this case

The evolution of the resonant angles 
{\bold{  $\Phi_{3,2,1} -\Phi_{2,1,2} = 3\lambda_3 -5\lambda_2 +2\lambda_1$, 
  $\Phi_{4,3,1} -\Phi_{3,2,2} = 3\lambda_4 -5\lambda_3 +2\lambda_2$ and   
  $\Phi_{5,4,1} -\Phi_{4,3,2} = 3\lambda_5 -5\lambda_4 +2\lambda_3$  }}
 are also shown in Fig.\ref{Fig11}.
 {\bold{These would exhibit small amplitude librations were there to be strict Laplace resonances.
 However, in these simulations the angles can take on values in the entire interval $(0,2\pi)$
 though the distributions for the first two show a small degree of nonuniformity.
 This is most pronounced for,  $\Phi_{4,3,1} -\Phi_{3,2,2}~=~3\lambda_4 -5\lambda_3 +2\lambda_2,$
 for which  the condition  (\ref{Lapstrict}) is closest to being satisfied initially (see Section \ref{sec20}).
 During the  simulations, fluctuations in the semi-major axes 
 (see Section {\ref {sec18}) lead to fluctuating departures from (\ref{Lapstrict})
 with amplitudes that significantly exceed
 the initial departures indicated in Section \ref{sec20}. Their amplitudes and persistence times enable the
 resonant  angles to range over $(0,2\pi)$ (see equation (\ref{Laplacec}).
 }}
 

The evolution of the semi-major axes for the six planets
is illustrated in  Fig.~\ref{Fig12} from which it can be seen that, 
after averaging out fluctuations, the innermost three
are moving inwards and the next two planets  $(j=4)$ and $(j=5)$  are moving outwards.
Any secular  movement of the outermost planet  is not expected as it is not in resonance
and it is seen to be significantly smaller.


For the innermost planet $(j=1)$ the value
 $d\log(a)/dt~=~-~~1.7\times~10^{-10}~y^{-1}$ 
was measured from the simulation, whereas the value 
 estimated  from (\ref{eqntid2snew}) after removing the variable scaling (and hence the parameters $\epsilon$ and $\lambda$) was
$d\log(a)/dt=-4.2 \times 10^{-10} y^{-1}.$

The corresponding values for planet $(j=2)$ were respectively 

\noindent $d\log(a)/dt= -1.2\times10^{-10} y^{-1}$
and $d\log(a)/dt= -8.7\times10^{-12} y^{-1}.$

\noindent For planet $(j=3)$ they were respectively $d\log(a)/dt= -3.3\times10^{-11} y^{-1}$
and 
 $d\log(a)/dt= 9.5\times10^{-11} y^{-1}.$
 
\noindent For planet $(j=4)$ they were respectively $d\log(a)/dt= 6.7\times10^{-11} y^{-1}$
and 
 $d\log(a)/dt= 1.9\times10^{-11} y^{-1}.$

\noindent For planet $(j=5)$ they  were respectively $d\log(a)/dt= 2.7\times10^{-10} y^{-1}$
and 

\noindent $d\log(a)/dt= 3.3\times10^{-12} y^{-1}.$

\noindent For planet $(j=6)$ no change could be measured fn the simulation
and as it was not in resonance we give no estimate from the semi-analytic model. 

These results reveal a discrepancy between the simulation and the semi-analytic model 
which is likely to be due to the presence of active Laplace resonances
as indicated in Section \ref{sec11}.
 According to the semi-analytic model, the dominant inward migration occurs for the innermost planet
 while the dominant outward migration occurs for  the planet (j=3). Others move significantly more slowly.
 However, in the simulation the innermost two planets move inward the most rapidly at comparable rates 
 while the planet $(j=3)$ moves inwards more slowly and
  planet (j=5) now moves outward the most rapidly. This indicates the interaction is spread among more planets
 than expected from the simple model because of linkage through the three Laplace resonances
 highlighted in Fig. \ref{Fig11}. The linking of more planets results in maximal migration rates  that  are somewhat  smaller.
{\bold{ In order to check the scaling of the above results with $Q',$ we have repeated the above simulation with $Q'=3.$
The evolution of the semi-major axes was  indeed  found to be consistent with being slowed down by a factor of three
as is the evolution of the resonant angles and eccentricities.}}
The most rapid inward migration  in the simulation occurred for the  innermost planet  {\bold{which, assuming this scales $\propto Q'$ occurs}
 on a time scale $\sim 2.5\times 10^{9}Q'  y.$ The time scale to significantly affect a period ratio
is around a hundred times less. Thus this will be significantly affected if  $ Q' <  100( t_a/ 2.5\times 10^{9}) y.$

\section{Discussion}\label{Disc}

In this paper we have developed  a  semi-analytic model for a planetary system consisting of a resonant chain  undergoing  orbital circularisation 
in Sections \ref{sec3} - \ref{sec7}.
This used  an approximation scheme
 which  assumed that near first order resonances among nearest neighbours dominated the dynamical interactions. 
  A set of variables  useful for  calculating the forced eccentricity response when changes in the semi-major axes  
could be neglected  was introduced in Section \ref{sec8} .
 In order to obtain conditions  enabling such an approximation,  scaled variables
were introduced in Section \ref{sec9}. The scaling
 involved  two small parameters, the first  characterising the typical ratio of planet mass to central mass, $\epsilon,$
and the  second,  $\epsilon^{2/3}/\lambda,$ with $\lambda$ assumed small but $<O(\epsilon^{2/3}),$
characterising the  magnitude of the deviation of the  near first order resonances from strict  commensurability. 
The calculation of the forced eccentricities  can be separated from consideration of the evolution of the 
semi-major axes, as was done in Section \ref{sec10}  when $\lambda$ is sufficiently small.    

Following this  procedure can be seen to be equivalent to calculating  forced eccentricities
 from the  epicyclic motion  produced  in response to
 perturbing planets assumed to be on on fixed circular orbits.
 This response can then used to calculate the rate of change of the semi-major axes.  
 In Section \ref{sec11} the possible presence of three body Laplace resonances
 was considered. When the conditions for these to occur are satisfied to a significantly  greater precision
 than the conditions for the first order resonances, features not included in the model
 are required to complete the procedure and determine the rate of change of the semi-major axes.
 That becomes unreliable if they are not included.
 
 The calculation of the forced eccentricities was described in Section \ref{sec12}
 and conditions for resonance angles to librate, together with the location ofthe 
centre of libration, 
 should that occur, was given in Section \ref{sec13}. Following on from this
 the calculation of the rate of change of the semi-major axes was given in Section \ref{sec14}.
 
 We then went on to perform numerical simulations of the
 HD 158259 and  EPIC 245950175 six planet systems in Section \ref{Numerics}.
 The aim was to determine  the effects of orbital circularisation as well as test the applicability of 
 the simple analytic model.
 
 In Section \ref{sec16} we gave a description of the parameters 
 of the HD 158259 system noting in Section \ref{sec17}  that the conditions for the
 occurrence of Laplace resonances are satisfied with approximately the
 same precision as the conditions for exact 3:2  first order commensurability  among these planets
 and so they are not expected, and indeed  not found to play a significant role. 
 Simulation results for $Q'=1,$ and $Q'=2$  were presented in Section \ref{sec18}. It was found that the simple
 analytic model was able to determine which resonant angles went into persistent libration
 and led to reasonable estimates of forced eccentricities in most cases.
 Furthermore the rate of evolution of the semi-major axes could also be reliably determined.
 Notably this system was found to be evolving towards a state in which two Laplace resonance
 conditions would be satisfied. 
 To avoid evolving significantly closer to strict Laplace resonances
  we estimated  that we need  $ Q' > 100 t_a/(2\times10^{9} y)$
with $t_a$ being the {\bold{time since formation}} in years. 
 
 We then went on to perform simulations of the 
  EPIC 245950175 system  giving a description of this in Section \ref{sec19}.
  As was noted in Section \ref{sec20}, in  contrast to the HD 158259 system,  the conditions for the
 occurrence of Laplace resonances are satisfied to much greater
  precision than are  the conditions for the first order 3:2 resonances, where these occur,
  and so they might be expected and indeed were found to play a significant role.
  
  Simulation results  for $Q'=1$ and $Q'=3$ were discussed in Section \ref{sec21}.
In this case the simple
 analytic model was also able to determine which resonant angles underwent  persistent libration
 and lead to reasonable estimates of forced eccentricities. However,
 the rate of evolution of the semi-major axes could not be determined reliably on account of
 the existence of Laplace resonances. These had the effect  of inducing comparable
 rates of change  amongst more planets at a somewhat reduced level.
 We found that in order for the deviation of a period ration from commensurability 
 not to be  significantly affected in the lifetime of the system  we needed
  $\sim Q' > \sim 100( t_a/ 2.5\times 10^{9}) y.$

 The above estimates indicate that tidal effects are likely to
 have significantly affected  some  aspects of the evolution of the systems if $Q'  < \sim 100$
 but not if $Q'$ significantly exceeds $\sim 10^3.$ In the latter case  the  active  Laplace resonances
 in the  EPIC 245950175 system would likely date back to formation as does the closeness to strict 
 3:2  commensurabilities in both systems. We remark that the forced eccentricities  in these systems are 
 typically $< 0.002,$  Thus accurate determinations of significantly larger values would rule out 
 the significance of orbital circularisation.
 
 An issue is the extrapolation of results obtained for low $Q'$ to much larger values
 made for numerical convenience.
{\bold {That the  evolution times should be $\propto Q',$ as found for the range of values we have considered,
  is in general expected for systems  where the evolution is driven by tides.
   It is also expected from consideration of the semi-analytic model developed in this paper.}}
 We have also checked that the relaxed states with librating resonant angles
   and  associated forced eccentricities also exist for much larger $Q'$ albeit for relatively
   short time scales and the applicability of the semi-analytic model  is reassuring.
   However, these aspects require further investigation.

\end{document}